\documentclass[12pt]{JHEP3}
\usepackage{youngtab}

\newcommand{\be}{\begin{equation}} 
\newcommand{\ee}{\end{equation}} 
\newcommand{\bea}{\begin{eqnarray}} 
\newcommand{\eea}{\end{eqnarray}} 
\def\eqa{&=&} 
\def\ccr{\nonumber\\} 
\def\ra{\rangle}

\title{Spinning particles and higher spin fields on (A)dS backgrounds}

\author{Fiorenzo Bastianelli $^{a,b}$,  Olindo Corradini $^{a,b}$  
and Emanuele Latini $^{a,c}$  \\  
\mbox{$^{a}$ Dipartimento  di Fisica, Universit{\`a} di Bologna, 
via Irnerio 46, I-40126 Bologna, Italy} \\ 
$^{b}$ INFN, Sezione di Bologna, via Irnerio 46, I-40126 Bologna, Italy \\ 
$^{c}$ INFN, Laboratori Nazionali di Frascati, CP 13, 
I-00044 Frascati, Italy\\ 

\mbox{E-mail: \email{bastianelli@bo.infn.it}, \email{corradini@bo.infn.it},
\email{latini@lnf.infn.it}}}

\abstract{Spinning particle models can be used to describe higher spin fields 
in first quantization. In this paper we discuss how spinning particles with 
gauged $O(N)$ supersymmetries on the worldline can be consistently coupled to 
conformally flat spacetimes, both at the classical and at the quantum level.
In particular, we consider canonical quantization on flat and on (A)dS 
backgrounds, and discuss in detail how the constraints due to the worldline
gauge symmetries produce geometrical equations for higher 
spin fields, i.e. equations written in terms of generalized curvatures. 
On flat space the algebra of constraints is linear, and one can integrate 
part of the constraints by introducing gauge potentials. This way
the equivalence of the geometrical formulation with the standard formulation 
in terms of gauge potentials is made manifest.
On (A)dS backgrounds the algebra of constraints becomes quadratic, 
nevertheless one can use it to extend much of the previous analysis to this case.
In particular, we derive general formulas for expressing the curvatures in 
terms of gauge potentials and discuss explicitly the cases of spin 2, 3 
and 4.}

\keywords{Supergravity models, Gauge symmetry, Sigma models}

\preprint{}

\begin{document}

\section{Introduction}
\label{sec:intro}

In a previous paper \cite{Bastianelli:2007pv} we have discussed the 
worldline quantization of massless higher spin fields, considering
in particular those fields that are described by spinning particle models 
with gauged $O(N)$ supersymmetries on the 
worldline \cite{Gershun:1979fb,Howe:1988ft,Howe:1989vn}
(which include all $D=4$ higher spin fields). 
We calculated the one-loop effective action in flat space, that
contains the information on the number of physical degrees 
of freedom propagating in the loop. This result was achieved by computing
the path integral of the $O(N)$ spinning particle on the circle.

To obtain more information on the quantum theory of higher spin fields
in a first quantized approach,
it is desirable to couple the spinning particles to more general backgrounds 
other than flat spacetime or, equivalently, to introduce suitable vertex 
operators to describe couplings to external particles. However,
this program has to face with the notorious difficulty of introducing 
interactions for higher spin fields\footnote{See for example 
\cite{Sorokin:2004ie} for a general introduction to the classical theory 
of higher spin fields, and 
\cite{Porrati:2008rm} which reviews and studies the problem of coupling 
spin 2 to higher spin particles in four dimensions
(see also \cite{Benincasa:2007xk} for a recent analysis).}.
This difficulty is evident also from the sigma model point 
of view. In fact, it was shown in \cite{Howe:1989vn} that for $N>2$
standard supersymmetry transformation rules leave the spinning particle 
action invariant only if the target spacetime is flat.  
The situation was improved in \cite{Kuzenko:1995mg}, 
where it was realized how to couple the spinning particle
to  maximally symmetric spaces, namely (A)dS spaces.
The construction presented in \cite{Kuzenko:1995mg} made use of the 
conformal invariance of the spinning particle which  was discovered
by Siegel, who embedded the model in a flat target space with two extra 
dimensions to keep conformal invariance manifest \cite{Siegel:1988ru} 
(this embedding had already been used by Marnelius for the case of $N=0,1$
\cite{Marnelius:1978fs}).

In this paper we perform a canonical analysis to study the couplings to 
curved spaces, and we are able to extend the known results to include couplings 
to arbitrary conformally flat spaces. 
This finding can be understood in a simple way:
noticing that the spinning particle action is invariant under a Weyl rescaling 
of the background target space metric is sufficient to guarantee 
consistent propagation on conformally flat manifolds.
The couplings to this class of curved spaces, even if mild, is presumably
not negligible, as one may expect some kind of conformal anomaly to give
rise to a nontrivial one loop effective action (more general than the 
one computed in \cite{Bastianelli:2007pv}). 
With this future application in mind, 
we proceed to study the canonical quantization of the model.
A canonical analysis is needed also to provide sufficient data for fixing the
counterterms that may arise when computing 
the corresponding path integral in curved spaces
\cite{Bastianelli:2006rx,Bastianelli:2005rc}, see in particular
\cite{Bastianelli:2002fv,Bastianelli:2002qw,Bastianelli:2005vk}
for the $N=0,1,2$ spinning particle cases, respectively.

Canonical quantization allows to identify the correct field equations
one is describing in first quantization. In the present case it
allows to make contact with the classical description
of higher spin fields in the so-called geometrical formulation,
dynamical equations originally 
proposed in \cite{Francia:2002aa1,Francia:2002aa2} 
which make use of the higher spin curvatures
constructed in \cite{Weinberg:1965rz,de Wit:1979pe}
(see \cite{Sorokin:2004ie} for reviews). 
This relation is seen by recalling that gauge symmetries
give rise to first class constraints that select physical states from
the Hilbert space. In flat space the constraints of the $O(N)$ spinning 
particle produce equations of motion written in terms 
of tensors that are interpreted as generalized curvatures describing higher 
spin fields. Gauge potentials can be introduced
by integrating a subset of these equations 
(those corresponding to the Bianchi identities).
This way one sees how the worldline approach reproduces and unifies 
various constructions that have appeared in the recent
literature on higher spin fields, like
the use of compensators to relax trace constraints
\cite{Francia:2002aa2,Sagnotti:2003qa} or
the use of generalized Poincar\'e lemmas to integrate
the Bianchi identities \cite{DuboisViolette:2001jk,Bekaert:2002dt,de Medeiros:2002ge,
Bandos:2005mb} and prove the equivalence with the standard formulation
of Fronsdal and Labastida \cite{Fronsdal:1978rb,Labastida:1986ft}
(see \cite{Sorokin:2004ie} for a list of references and 
discussions of related works). 
We present the analysis in arbitrary dimensions $D$, but 
only for the case of even $N$, i.e. for particles with integer
spin $s =\frac{N}{2}$. Extension to the odd $N$ case
should proceed in a similar fashion.

Then we analyze the constraint equations in the case of (A)dS spaces. 
The algebra of constraints is again first class, 
but the algebra closes only quadratically. 
It is interesting to note that
this algebra coincides with the zero mode sector of 
the Bershadsky-Knizhnik $SO(N)$-extended superconformal 
algebra in two dimensions \cite{Bershadsky:1986ms,Knizhnik:1986wc}.
The constraints produce again geometrical equations of motion for the higher spin 
curvatures on (A)dS spaces. 
Quadratic closure complicates the algebraic structure, which nevertheless
remains of valuable help. In fact, we use it to express the curvatures 
in terms of higher spin gauge potentials. Then, we consider in detail
the cases of spin $s=2,3,4$, with the $s=2$ case
corresponding to the familiar case of the graviton if $D=4$.
Quadratic algebras have appeared before in the description of higher spin fields,
see for example \cite{Buchbinder:2001bs,Sagnotti:2003qa}.

Though not discussed in this paper, one may find in the literature 
other particle models related to higher
spin fields, like the twistor-like particle of refs. \cite{Bandos:1998vz,Bandos:1999qf,Bandos:2005mb}
or particles that could be constructed 
using the $OSp$ quantum mechanics of ref. \cite{Hallowell:2007qk}.
The same BRST approach of refs. \cite{Pashnev:1997rm}
used to describe higher spin field equations can perhaps be
related to a particle model.

In the following we shall structure our paper as indicated 
in the table of content.

\section{The $O(N)$ spinning particle} 

In this section we first review the classical formulation
of the spinning particle propagating in Minkowski space. Then,
we proceed to describe the coupling to conformally flat spaces.

\subsection{Minkowski space}

It will be useful to present the  $O(N)$ spinning particle 
action directly in phase space. 
The dynamical variables are given by: the cartesian coordinates  $x^\mu$
of the particle moving in a $D$ dimensional Minkowski space, 
their conjugate momenta $p_\mu$, and $N$ real Grassmann variables 
with spacetime vector indices $\psi_i^\mu $ ($i =1,..,N$).
The Minkowski metric $\eta_{\mu\nu}\sim (-,+,\cdots,+)$ is 
used to raise and lower spacetime indices.
In addition, there is an $O(N)$-extended supergravity on the worldline, 
whose gauge fields are given by the einbein $e$, the gravitinos $\chi_i$, and 
the $SO(N)$ gauge field  $a_{ij}$.
The action which defines the model is given by
\bea
S \eqa \int dt 
\Big [ p_\mu \dot x^\mu + {i \over 2} \psi_{i\mu} \dot \psi_i^\mu 
-e \underbrace{\Big ( {1\over 2} p_\mu p^\mu \Big )}_{H}
- i \chi_i \underbrace{\Big ( p_\mu \psi^\mu_i\Big )}_{Q_i} 
- {1 \over 2} a_{ij} \underbrace{\Big (i \psi_i^\mu \psi_{j\mu} 
\Big )}_{J_{ij}}
\Big ]  
\label{action1}
\eea
where $H, Q_i, J_{ij}$ denote the first class constraints gauged 
by the fields $e,\chi_i,a_{ij}$.
The kinetic term defines the phase space symplectic form
and fixes the graded Poisson brackets:
$\{x^\mu, p_\nu \}_{_{PB}}=\delta^\mu_\nu$ and  
$\{\psi_i^\mu,\psi_j^\nu\}_{_{PB}} = -i \eta^{\mu\nu} \delta_{ij}$.
With these brackets one can easily compute the constraint algebra
at the classical level
\bea
\{Q_i,Q_j\}_{_{PB}} \eqa -2i \delta_{ij} H \ccr[1mm]
\{J_{ij},Q_k\}_{_{PB}} \eqa \delta_{jk} Q_i -\delta_{ik} Q_j  \ccr[1mm]
 \{J_{ij},J_{kl}\}_{_{PB}} \eqa \delta_{jk} J_{il} - \delta_{ik} J_{jl} 
- \delta_{jl} J_{ik} + \delta_{il} J_{jk}  
\label{linear}
\eea
which is first class and thus gauged consistently by the fields  
$e,\chi_i,a_{ij}$. 
This algebra
is known as the {\em $O(N)$-extended susy algebra}: it has $N$ susy charges  
$Q_i$ which close on the Hamiltonian $H$ and which transform in the vector 
representation of $SO(N)$, whose Lie algebra 
is described by the last line. We now discuss the various 
symmetries of the model.

The gauge symmetries are those of the $O(N)$-extended supergravity on 
the worldline, whose infinitesimal gauge transformations with parameters 
$\xi,\epsilon_i,\alpha_{ij}$ are given by 
\bea
\delta x^\mu \eqa \{ x^\mu , G \}_{_{PB}} = 
\xi p^\mu + i\epsilon_i \psi^\mu_i  \ccr
\delta p_\mu \eqa \{ p_\mu , G \}_{_{PB}} = 0 \ccr
\delta \psi_i^\mu \eqa \{ \psi_i^\mu , G \}_{_{PB}} 
= -\epsilon_i p^\mu + \alpha_{ij} \psi_j^\mu 
\ccr
\delta e \eqa \dot \xi + 2 i \chi_i \epsilon_i \ccr
\delta \chi_i \eqa \dot \epsilon_i  - a_{ij}\epsilon_j 
+\alpha_{ij} \chi_j \ccr
\delta a_{ij} \eqa \dot \alpha_{ij} +\alpha_{im} a_{mj} +
\alpha_{jm} a_{im}  
\label{gauge-tr}
\eea
where $G\equiv \xi H + i\epsilon_i Q_i + {1\over 2} \alpha_{ij} J_{ij}$
denotes the generator of gauge transformations. One could add trivial symmetries
proportional to the equations of motion to present the worldline diffeomorphisms
in the standard geometrical form, but this is not so natural in the hamiltonian
formalism.

The rigid symmetries include transformations under 
the Poincar\'e group of target space, which guarantees the
relativistic invariance of model. They are given by
\bea
\delta x^\mu = \omega^{\mu}{}_\nu x^\nu + a^\mu\ , \quad
\delta p_\mu = \omega_{\mu}{}^\nu p_\nu \ , \quad
\delta \psi_i^\mu = \omega^{\mu}{}_\nu  \psi_i^\nu 
\eea
where $\omega^{\mu}{}_\nu$ and $a^\mu$ specify infinitesimal Lorentz 
rotations and spacetime translations, respectively.
The worldline gauge fields are left invariant by these symmetries.

In addition, the model is conformal invariant.
To prove this we first show that the model has background 
symmetries\footnote{These are symmetries in which
also the background fields, like the spacetime metric, transform.} 
corresponding to: $(i)$ diffeomorphisms,  $(ii)$ local Lorentz 
transformations, $(iii)$ Weyl rescalings of the flat target space metric.
Then, conformal Killing vectors, which by definition leave invariant 
the background metric, identify rigid symmetries of the model.
They generate the conformal group $SO(D,2)$.

To discuss these background symmetries we find it convenient to rewrite the 
action (\ref{action1}) using arbitrary coordinates, denoted again by 
$x^\mu$. We also denote the Minkowski metric in arbitrary coordinates 
by $g_{\mu\nu}$. Then
we introduce an orthonormal tangent frame specified by the vielbein 
$e_\mu{}^a$ and use  $\psi^a_i\equiv \psi^\mu_ie_\mu{}^a(x)$ as 
independent variables.
Given the vielbein one may construct the unique spin connection 
$\omega_{\mu ab}$, which enters the definition of the covariant momenta
\be
\pi_\mu=p_\mu - \frac{i}{2}\omega_{\mu ab} \psi^a_i \psi^b_i  \ .
\ee
The coefficient in front of the spin connection is easily fixed by requiring 
the covariance condition
\be
\{ \pi_\mu , \pi_\nu \}_{_{PB}} = \frac{i}{2} R_{\mu\nu ab} \psi^a_i \psi^b_i  
\ee
so that in flat space the covariant momenta commute. 
With  these tools at hand
the action (\ref{action1}) can be rewritten in the form
\bea
S \eqa \int dt 
\Big [ p_\mu \dot x^\mu + {i \over 2} \psi_{i a} \dot \psi_i^a 
-e \underbrace{\Big ( {1\over 2} g^{\mu\nu}\pi_\mu \pi_\nu \Big )}_{H}
- i \chi_i \underbrace{\Big ( \psi^a_i e_a{}^\mu \pi_\mu \Big )}_{Q_i} 
- {1 \over 2} a_{ij} \underbrace{\Big (i \psi_i^a \psi_{j a} \Big )}_{J_{ij}}
\Big ]  \ . \qquad 
\label{action2}
\eea
We are now ready to discuss its background symmetries:

$(i)$ Diffeomorphisms of target space are identified quite easily.
The coordinates transform as usual, $x^\mu \to {x^\mu}'(x)$,
the momenta as a 1-form, 
$p_\mu \to {p_\mu}'= p_\nu \frac{\partial x^\nu}{{\partial x^\mu}'}$, 
and the background fields $g_{\mu\nu}, e_\mu{}^a, \omega_{\mu ab}$ 
as tensors as indicated by their coordinate indices.
The fermions $\psi^a_i$ are left invariant, just like 
the supergravity gauge fields $e, \chi_i,a_{ij}$.
These transformations are easily seen to be an invariance of the action.

$(ii)$ Proving local Lorentz invariance is slightly more difficult.
An infinitesimal local Lorentz transformation is specified by 
the parameters $\lambda^{ab}(x)=-\lambda^{ba}(x)$.
It leaves the coordinates $x^\mu$ invariant and transforms 
the worldline fermions as vectors 
\be 
\delta \psi^a_i=\lambda^a{}_b(x)\psi^b_i \ .
\ee
The symplectic term of the action is left invariant if one assigns to 
the momenta the transformation rule
\be 
\delta p_\mu= - \frac{i}{2}\partial_\mu \lambda_{ab}(x)\psi^a_i \psi^b_i   \ .
\ee
The background fields $g_{\mu\nu}, e_\mu{}^a, \omega_{\mu ab}$ 
transform as usual under local Lorentz transformations, 
and in particular the spin connection transforms as the local Lorentz
gauge field 
\be 
\delta \omega_\mu{}^{ab}= - \partial_\mu \lambda^{ab}+
\lambda^a{}_c\, \omega_\mu{}^{cb}+ \lambda^b{}_c\, \omega_\mu{}^{ac} \ .
\ee
As a consequence the covariant momentum $\pi_\mu$ is left invariant.
Therefore the full action is invariant.

$(iii)$ Finally, let us prove invariance under Weyl rescalings of the
target space metric.
Under an infinitesimal Weyl rescaling specified by the local 
parameter $\phi(x)$, 
which is a function of target space, the background fields transform as
\bea
\delta g_{\mu\nu} = 2 \phi\, g_{\mu\nu} \ , \quad
\delta  e_\mu{}^a = \phi\, e_\mu{}^a \ , \quad
\delta \omega_\mu{}^{ab}= (e_\mu{}^a e_\nu{}^b-e_\mu{}^b e_\nu{}^a)
\nabla^\nu \phi \ .
\eea
As a consequence the covariant momentum transforms
as 
\be
\delta \pi_\mu = -i \psi_{\mu i}  \psi^\nu_i \partial_\nu \phi 
\ee
and the constraints as
\bea
\delta Q_i \eqa -\phi\, Q_i - J_{ij}\psi^\mu_j \partial_\mu \phi 
\ccr
\delta  H \eqa -2 \phi\, H + i \psi^\mu_i \partial_\mu \phi\, Q_i \ .
\eea
These transformations can be compensated by suitable
transformations on the worldline gauge fields
\bea
\delta e \eqa  2 \phi\, e \ccr 
\delta \chi_i \eqa -e \psi_i^\mu \partial_\mu \phi + \chi_i \phi  \ccr 
\delta a_{ij} \eqa i (\chi_i \psi_j^\mu -\chi_j \psi_i^\mu)
\partial_\mu \phi  
\eea
while the variables $x^\mu, p_\mu, \psi^a_i$ are taken to be invariant.
This proves Weyl invariance.

Because of these background symmetries,
conformal Killing vectors necessarily produce global symmetries. In fact,
the conformal Killing vectors are precisely those vector fields $\xi^\mu$
that generate 
infinitesimal diffeomorphisms whose effect on the metric and on the vielbein
can be compensated by suitable Weyl and local Lorentz transformations,
\bea
\delta g_{\mu\nu} \eqa  {\cal L}_\xi g_{\mu\nu} + 2\phi g_{\mu\nu} =0 \ccr
\delta e_\mu{}^a  \eqa  {\cal L}_\xi e_\mu{}^a + \phi e_\mu{}^a  +  
\lambda^a{}_b e_\mu{}^b=0 
\eea
where ${\cal L}_\xi$ denotes the Lie derivative acting along the 
vector field $\xi^\mu$. 
As the background fields are left untransformed, the conformal Killing 
vectors induce rigid symmetries of the action (\ref{action2}).
They generate the conformal group  $SO(D,2)$, 
which extend the Poincar\'e group to include 
scale transformations and conformal boosts.

An additional bonus of the background Weyl symmetry is that it guarantees
that the $O(N)$ spinning particle propagates consistently on arbitrary 
conformally flat manifolds. These spaces include the class of 
maximally symmetric spaces, i.e. the (A)dS spaces, 
which were shown to be consistent backgrounds 
for the spinning particle in \cite{Kuzenko:1995mg}, but are more general.

Before closing, let us report the finite Weyl transformations
leaving the action invariant. They are given by
\bea
g_{\mu\nu}' = {\rm e}^{2 \phi} g_{\mu\nu} \ , \quad
{e_\mu{}^a}' = {\rm e}^{\phi} e_\mu{}^a \ , \quad
{\omega_\mu{}^{ab}}'= \omega_\mu{}^{ab}+
(e_\mu{}^a e_\nu{}^b -e_\mu{}^b e_\nu{}^a) \nabla^\nu \phi  \ ,
\eea
implying
\bea
{Q_i}' \eqa {\rm e}^{-\phi} 
\Big( Q_i -J_{ij}\psi_j^\mu \partial_\mu \phi\Big)
\ccr
H' \eqa {\rm e}^{-2 \phi}  \Big ( H - i Q_i \psi_i^\mu \partial_\mu \phi
-\frac{i}{2} J_{ij} \psi_i^\mu \partial_\mu \phi
\psi_j^\nu \partial_\nu \phi \Big) \ ,
\eea
and 
\bea
e' \eqa  {\rm e}^{2 \phi} e
\ccr 
{\chi_i}' \eqa {\rm e}^{\phi}\Big ( \chi_i - 
e \psi^\mu_i\partial_\mu \phi \Big)
\ccr 
{a_{ij}}' \eqa a_{ij} + i(\chi_i \psi_j^\mu -\chi_j \psi_i^\mu)
\partial_\mu \phi
 -i e \psi^\mu_i\partial_\mu \phi \psi^\nu_j\partial_\nu \phi  \ .
\eea

\subsection{Conformally flat spaces}
\label{sec:2.2}

As just discussed, the background Weyl symmetry implies that the spinning particle 
is consistent on any conformally flat spacetime.
In this section we verify this claim by direct canonical analysis.

The form of the action is the same as the one reported in eq. (\ref{action2})
\bea
S \eqa \int dt 
\Big [ p_\mu \dot x^\mu + {i \over 2} \psi_{i a} \dot \psi_i^a 
-e \underbrace{\Big ( {1\over 2} g^{\mu\nu}\pi_\mu \pi_\nu \Big )}_{H_0}
- i \chi_i \underbrace{\Big ( \psi^a_i e_a{}^\mu \pi_\mu \Big )}_{Q_i} 
- {1 \over 2} a_{ij} \underbrace{\Big (i \psi_i^a \psi_{j a} \Big )}_{J_{ij}}
\Big ]   \qquad
\label{action3}
\eea
but we have renamed the hamiltonian as $H_0$ in view of convenient 
redefinitions to be introduced later.
We will start assuming an arbitrary metric $g_{\mu\nu}$, and 
verify that the constraints $H_0 , Q_i ,J_{ij} $
continue to form a first class algebra on spaces that are conformally flat, 
so that by assigning suitable transformation rules to the gauge fields 
$e, \chi_i, a_{ij}$ the action keeps on being gauge invariant.

As anticipated, it is instructive to begin by considering generic curved spaces.
Apart from the $SO(N)$ subalgebra generated by the $J_{ij}$,
which remains unmodified, one obtains the following algebra
\bea
\{Q_i,Q_j\}_{_{PB}} \eqa -2i \delta_{ij} H_0
 +\frac{i}{2} R_{abcd} \psi^a_i \psi^b_j  \psi^c \cdot \psi^d \ccr
\{Q_i,H_0\}_{_{PB}} \eqa 
-\frac{i}{2} \pi^a R_{abcd} \psi^b_i \psi^c \cdot \psi^d 
\label{alg1}
\eea
which generically fails to be first class. Of course, one could try to add
new constraints to force the algebra to close, 
but this may overconstrain the system.

An option, that in the light of the previous analysis
is guaranteed to work, is to restrict attention to conformally flat spaces.
These spaces have a vanishing Weyl tensor, which allows to solve 
the Riemann tensor in terms of the Ricci tensor and curvature scalar
\bea
R_{abcd}\eqa 
\frac{1}{(D-2)} \Big ( \eta_{ac}R_{bd} -\eta_{ad}R_{bc} -\eta_{bc}R_{ad}+ 
\eta_{bd}R_{ac} \Big )
\ccr
&& - \frac{R}{(D-2)(D-1)} 
\Big ( \eta_{ac}\eta_{bd} -\eta_{ad}\eta_{bc} \Big )  \ .
\label{cfriemann}
\eea
Substituting this relation into (\ref{alg1}) produces
\bea
\{Q_i,Q_j\}_{_{PB}} \eqa -2i \delta_{ij} H_0
-\frac{iR}{(D-2)(D-1)} J_{ik}J_{jk}
-\frac{R_{ab}}{(D-2)} \Big (\psi^a_i \psi^b_k J_{jk} +
(i \leftrightarrow j)\Big )
\ccr
\{Q_i,H_0\}_{_{PB}} \eqa \frac{R}{(D-2)(D-1)} Q_k J_{ki} 
+ \frac{R_{ab}}{(D-2)} \Big (\pi^a \psi^b_k J_{ik} + i 
\psi^a_i \psi^b_k Q_k \Big ) 
\label{alg2}
\eea
which 
becomes first class, 
though with structure functions rather than structure constants.
This is enough to guarantee consistency of the gauge system
at the classical level, see for example \cite{Henneaux:1985kr}.

It may be convenient, especially when considering maximally symmetric spaces,
to redefine the hamiltonian as
\be
H = H_0+\Delta H= \frac{1}{2} g^{\mu\nu}\pi_\mu \pi_\nu
-\frac{1}{8}R_{abcd}\psi^a \cdot \psi^b \psi^c \cdot \psi^d
\label{2.23}
\ee
so that on general curved spaces the algebra (\ref{alg1})
takes the form
\bea
\{Q_i,Q_j\}_{_{PB}} \eqa -2i \delta_{ij} H
+\frac{i}{2} R_{abcd} \Big ( \psi^a_i \psi^b_j  -\frac{1}{2}\delta_{ij} 
\psi^a \cdot \psi^b \Big ) \psi^c \cdot \psi^d 
\ccr
\{Q_i,H \}_{_{PB}} \eqa 
\frac{1}{8}  \psi^e_i \nabla_e R_{abcd} 
\psi^a\cdot \psi^b \psi^c \cdot \psi^d  \ .
\label{alg3}
\eea
Written in this way one sees that 
the second Poisson bracket vanishes on locally 
symmetric spaces, but the first one remains second class. 
Thus, the model is inconsistent on generic curved spaces for $N>2$ 
(while for $N \leq 2$ one can show that the offending terms vanish).
On conformally flat spaces these relations simplify to 
 \bea
\{Q_i,Q_j\}_{_{PB}} \eqa -2i \delta_{ij} H
 +\frac{iR}{(D-2)(D-1)} \Big (\frac{1}{2}\delta_{ij} J_{kl} J_{kl} 
-J_{ik} J_{jk}\Big ) \\
&& -\frac{R_{ab}}{(D-2)} \Big (\psi^a_i \psi^b_k J_{jk}
+\psi^a_j \psi^b_k J_{ik} -\delta_{ij} \psi^a_k \psi^b_l J_{kl} \Big )
\ccr
\{Q_i,H \}_{_{PB}} \eqa 
-\frac{1}{4(D-2)(D-1)} \psi^c_i \nabla_c R J_{kl} J_{kl} 
+\frac{i}{2(D-2)} \psi^c_i \nabla_c R_{ab} \psi^a_k \psi^b_l J_{kl} 
\nonumber
\label{alg4}
\eea
with 
\bea
H = H_0 
+\frac{R}{4(D-2)(D-1)} J_{ij} J_{ij} -\frac{i R_{ab}}{2(D-2)} \psi^a_i \psi^b_j 
J_{ij} \ .
\label{2.25}
\eea
The corresponding action on conformally flat spaces
\bea
S \eqa \int dt 
\Big [ p_\mu \dot x^\mu + {i \over 2} \psi_{i a} \dot \psi_i^a 
-e H - i \chi_i Q_i - {1 \over 2} a_{ij} J_{ij}
\Big ]   \qquad
\label{action4}
\eea
is then gauge invariant under suitable transformation rules generated by 
the constraints and their structure functions.
We refrain from presenting them here.

All these expressions simplify further on maximally symmetric
spaces, the (A)dS spaces, which are a subset of conformally flat spaces.
As we are going to treat the canonical quantization 
of these cases in some detail,
it may be useful to report the corresponding classical formulas.
The Riemann tensor for maximally symmetric spaces is of the form
\be
R_{abcd}= b (\eta_{ac}\eta_{bd}-\eta_{ad}\eta_{bc})
\ee
where the constant $b$ is related to the curvature scalar by
$b=\frac{R}{D(D-1)}$. 
The improved hamiltonian now reads as
\be 
H=H_0+\Delta H = \frac{1}{2} \pi^a \pi_a -\frac{b}{4}J_{ij}J_{ij}
\ee
and the complete gauge algebra, including the $J_{ij}$ charges, has the following
nonvanishing Poisson brackets
\bea
\{Q_i,Q_j\}_{_{PB}} \eqa -2i \delta_{ij} H +ib \Big (J_{ik}J_{jk}
-\frac{1}{2}\delta_{ij} J_{kl} J_{kl}\Big) \ccr[1mm]
\{J_{ij},Q_k\}_{_{PB}} \eqa \delta_{jk} Q_i -\delta_{ik} Q_j  \ccr[1mm]
 \{J_{ij},J_{kl}\}_{_{PB}} \eqa \delta_{jk} J_{il} - \delta_{ik} J_{jl} 
- \delta_{jl} J_{ik} + \delta_{il} J_{jk} \ .
\label{nonlinear}
\eea
It is a quadratic deformation of the linear algebra in (\ref{linear}),
with $b$ playing the role of deforming parameter.
It is interesting to note that this algebra reproduces the 
(classical version) of the zero mode sector of certain two-dimensional
nonlinear superconformal algebras introduced some time ago by
Bershadsky and Knizhnik \cite{Bershadsky:1986ms,Knizhnik:1986wc}.
The corresponding action (\ref{action4}) is invariant under
transformation rules that can be easily derived using 
the constraints and their structure functions.
We list them here, as they might be useful in discussing gauge fixing issues
\bea
\delta x^\mu \eqa \{ x^\mu , G \}_{_{PB}} = 
\xi \pi^\mu + i\epsilon_i \psi^\mu_i  \ccr
\delta p_\mu \eqa \{ p_\mu , G \}_{_{PB}} =
(\xi \pi^a + i\epsilon_k\psi^a_k) \Big (\frac{i}{2} \partial_\mu \omega_{abc} 
\psi^b_i \psi^c_i -p_\nu \partial_\mu e_a{}^\nu \Big )
\ccr
\delta \psi_i^a\eqa \{ \psi_i^a , G \}_{_{PB}} 
= -(\xi \pi^b +i \epsilon_k \psi_k^b) \omega_{bac}\psi_i^c
-\epsilon_i \pi^a + (\alpha_{ij}-\xi b J_{ij}) \psi_j^a 
\ccr
\delta e \eqa \dot \xi + 2 i \chi_i \epsilon_i 
\ccr
\delta \chi_i \eqa \dot \epsilon_i  - a_{ij}\epsilon_j +\alpha_{ij} \chi_j 
\ccr
\delta a_{ij} \eqa \dot \alpha_{ij} +\alpha_{im} a_{mj} + \alpha_{jm} a_{im}
+ib \Big (\chi_k \epsilon_k J_{ij} 
+\sigma (\epsilon_i \chi_k J_{kj} - \epsilon_j \chi_k J_{ki})
\ccr
&&
+(1-\sigma)  (\epsilon_k J_{kj}\chi_i -\epsilon_k J_{ki}\chi_j)\Big )
\label{gauge-tr1}
\eea
where the free parameter $\sigma\in[0,1]$ labels different choices of 
splitting the algebra in structure functions and generators.

This hamiltonian formulation of the spinning particle on (A)dS spaces
is equivalent to the lagrangian formulation discussed by
Kuzenko and Yarevskaya in \cite{Kuzenko:1995mg}.

\section{Canonical quantization} 

In this section we study canonical quantization of the spinning particle
on the class of spaces just discussed. Phase space variables become operators
and the problem is to find the correct ordering 
that preserves the first class property of the constraints. 
As we shall discuss, this requirement introduces quantum corrections to the 
classical hamiltonian as well.
The quantum constraint equations are then used to select the physical sector 
of the Hilbert space, and are interpreted as field equations for higher 
spin fields.

\subsection{Minkowski space}

Let us briefly review canonical quantization for the $O(N)$ spinning particle 
in flat space, which is best carried out using cartesian coordinates.
The fundamental (anti) commutation relations are obtained
from the corresponding classical Poisson brackets
and read (from now on all variables are operators)
\be
[x^\mu, p_\nu] = i \delta^\mu_\nu \ , \qquad 
\{\psi_i^\mu,\psi_j^\nu\} = \eta^{\mu\nu} \delta_{ij} \ .
\ee
This operator algebra is realized irreducibly 
on a Hilbert space which contains also unphysical states. 
The physical states are obtained \`a la Dirac-Gupta-Bleuler by
requiring the constraints to annihilate them.
Of course, the quantum constraints are constructed from the classical
ones by specifying a suitable ordering plus possible quantum corrections.
In the case of flat spacetime, 
one only needs to specify the correct ordering in the definition of 
the $SO(N)$ generators, as there are no other ordering ambiguities.
Taking that into account, the quantum constraint are given by
\bea
H = {1\over 2} p_\mu p^\mu \ ,\qquad
Q_i = p_\mu \psi_i^\mu \ , \qquad
J_{ij} = \frac{i}{2} [\psi_i^\mu, \psi_{j\mu}]  
\label{cons-flatsp}
\eea
and satisfy the quantum algebra
\bea
\{Q_i,Q_j\} \eqa 2 \delta_{ij} H 
\label{3.3}
\\[1mm]
[J_{ij},Q_k] \eqa i\delta_{jk} Q_i -i \delta_{ik} Q_j  
\label{3.4}
\\[1mm]
[J_{ij},J_{kl}] \eqa i\delta_{jk} J_{il} - i\delta_{ik} J_{jl} 
- i\delta_{jl} J_{ik} + i\delta_{il} J_{jk} 
\label{3.5}
\eea
which is first class.
The corresponding constraints give rise to higher spin field 
equations \cite{Gershun:1979fb,Howe:1988ft,Howe:1989vn},
in the form originally developed by Bargmann and Wigner.
These equations are described by a multispinor $\Psi_{\alpha_1,..,\alpha_N}$ 
that satisfies a Dirac equation in each index and, 
in addition, suitable algebraic constraints which project 
onto the irreducible spin $\frac{N}{2}$ components \cite{Bargmann:1948ck}.
We shall discuss these equations in a different basis for the case of even $N$ 
(integer spin) in section \ref{sec4}. 
The alternative BRST quantization for this model is described 
in refs. \cite{Marnelius:1988ab} and \cite{Siegel:1999ew}.
In particular in \cite{Siegel:1999ew} one finds its use to construct 
second quantized actions for any spin in flat spaces of arbitrary dimensions.

\subsection{Conformally flat spaces}

The classical structure presented in section \ref{sec:2.2}
carries over to the quantum theory
after specifying the correct orderings that preserve the symmetries 
of the model.
It is again useful to discuss first the case of generic curved spaces, 
and then restrict to conformally flat spaces which will be shown
to admit a first class constraint algebra.

The quantum algebra of the fundamental operators now reads as
\be
[x^\mu, p_\nu] = i \delta^\mu_\nu \ , \qquad 
\{\psi_i^a,\psi_j^b\} = \eta^{ab} \delta_{ij} 
\ee
since worldline fermions with flat indices are taken as fundamental variables.
The correct ordering of the $SO(N)$ currents is again immediate
\be
J_{ij} = \frac{i}{2} [\psi_i^a, \psi_{j a}] \ .
\ee 
The susy charges are also ordered uniquely as follows\footnote{
For notational simplicity we use nonhermitian operators $Q_i$. 
Hermiticity is obtained by a similarity transformation 
$A \to g^{\frac{1}{4}} A g^{-\frac{1}{4}}$ on the quantum variables, 
so that hermitian operators $Q_i$ (as well as $H$) are obtained by substituting
$p_\mu \to g^{\frac{1}{4}}p_\mu g^{-\frac{1}{4}}$,
see for example \cite{Bastianelli:2006rx}.}
\be
Q_i = \psi_i^a e_a{}^\mu 
\Big ( p_\mu - \frac{i}{2}\omega_\mu{}^{bc} \psi_j^b  \psi_j^c \Big ) \ .
\ee 
To understand why this covariantization is unique, one may recall
that it corresponds to the unique covariant derivative acting on a 
multispinorial wave function.

Before proceeding, it may be useful to introduce the hermitian 
Lorentz generators 
\be 
M^{ab} = {i\over 2} [\psi_j^a,  \psi_j^b]
\ee
which satisfy the Lorentz algebra and commute with the $SO(N)$ generators
\bea
[M^{ab},M^{cd}] \eqa
 i \eta^{bc} M^{ad} -i \eta^{bd} M^{ac} -i \eta^{ac} M^{bd} +i \eta^{ad} M^{bc} 
\ccr[1mm]
[M^{ab},J_{ij}] \eqa 0 \ .
\eea
Then one can write the covariant momentum in the form
$\pi_\mu =p_\mu - \frac{1}{2}\omega_{\mu ab}  M^{ab}$
and the susy charges as
$ Q_i = \psi_i^a e_a{}^\mu  \pi_\mu = \psi_i^a \pi_a $.

At this point one may start checking the algebra on generic curved 
spaces and identify a suitable hamiltonian operator. 
Equations (\ref{3.4}) and (\ref{3.5}) are left unmodified, but the other
(anti)commutators produce
\bea
\{Q_i,Q_j\} \eqa 2 \delta_{ij} H_0  
- \frac{1}{2} \psi_i^a \psi_j^b R_{abcd} M^{cd} \\[1mm]
[Q_i,H_0] \eqa \frac{1}{2} R_{ab} \psi^a_i \pi^b
+\frac{i}{2}R_{abcd}\psi_i^a M^{cd}\pi^b
-\frac{1}{2}\nabla_a R_{bc}\psi_i^c M^{ab}
\label{uno}
\eea
where
\be
H_0= \frac{1}{2} \Big (\pi^a \pi_a -i \omega^a{}_{ab} \pi^b \Big) 
\label{ho}
\ee
corresponds to the minimal quantum covariantization
of the classical operator appearing in (\ref{action3}): in particular,
the second term in $H_0$ is a quantum correction which guarantees covariance.
As in the classical case, also in the quantum case the algebra fails to be 
first class, implying a generic inconsistency on arbitrary spaces.

Thus, we restrict to conformally flat spaces. Using the relation 
(\ref{cfriemann}) for the Riemann tensor on conformally flat spaces,
we obtain the quantum version of (\ref{alg2}) which takes the form
\bea
\{Q_i,Q_j\} \eqa 2\delta_{ij} H -\frac{i}{(D-2)} R_{ab} 
\Big ( \psi^a_i \psi^b_k J_{jk} + \psi^a_j \psi^b_k J_{ik}
-\delta_{ij} \psi^a_k \psi^b_l J_{kl}\Big)
\ccr[1mm]
&+& \frac{1}{2(D-1)(D-2)}R \Big ( J_{ik}J_{jk}+J_{jk}J_{ik}
-\delta_{ij} J_{kl}J_{kl} \Big)
\ccr[1mm]
[Q_i,H] \eqa \frac{1}{4(D-1)}\nabla_a R\, \psi^a_k J_{ik}
-\frac{i}{4(D-1)(D-2)}\nabla_a R\, \psi^a_i
J_{jk}J_{jk}\ccr[1mm]
&-& \frac1{2(D-2)} \nabla_a R_{bc}\, \psi^a_i \psi^b_j\psi^c_k\ J_{jk}
\eea
where
\bea
H &=& H_0 +\frac{1}{8} R_{abcd} M^{ab} M^{cd} 
-\frac{ (N-2)(D+N-2)}{16 (D-1)}\, R
\\[1mm]
&=& H_0 +\frac{1}{4(D-1)(D-2)}\, R J_{jk}J_{jk}
-\frac{i}{2(D-2)} R_{ab}\, \psi^a_j\psi^b_k\, J_{jk}
+\frac{(D+N-2)}{8(D-1)}\, R
\nonumber
\eea
with $H_0$ as in (\ref{ho}).
The result is that, with a suitable quantum redefinition of the hamiltonian 
$H$, the algebra closes and becomes first class.
The last term in both expressions of $H$, proportional to the scalar curvature, 
is a quantum effect that did not appear in the corresponding
classical expressions (\ref{2.23}) and (\ref{2.25}).
This final result proves the quantum consistency  of the model on conformally 
flat spaces.

\subsection{(A)dS spaces}

The subset of maximally symmetric spaces, characterized by  
a Riemann tensor of the form
$ R_{abcd}=b(\eta_{ac}\eta_{bd}- \eta_{ad}\eta_{bc})$,
is much simpler. In fact, the above algebra simplifies further and we 
summarize here the set of quantum constraints appropriate for (A)dS spaces
\bea
J_{ij} \eqa \frac{i}{2} [\psi_i^a, \psi_{j a}]  \ccr
Q_i \eqa \psi_i^a e_a{}^\mu 
\Big ( p_\mu - \frac{1}{2}\omega_{\mu bc} M^{bc}\Big )
\ccr
H \eqa
\frac{1}{2} \Big (\pi^a \pi_a -i \omega^a{}_{ab} \pi^b \Big)
-\frac{b}{4} J_{ij}J_{ij} -b A(D)
\label{17}
\eea
where $A(D) \equiv (2-N)\frac{D}{8}- \frac{D^2}{8}$, and the 
corresponding quantum algebra 
\bea
[J_{ij},J_{kl}] \eqa i\delta_{jk} J_{il} - i\delta_{ik} J_{jl} 
- i\delta_{jl} J_{ik} + i\delta_{il} J_{jk} \ccr[1mm] 
[J_{ij},Q_k]\eqa i\delta_{jk} Q_i -i \delta_{ik} Q_j  \ccr[1mm]
\{Q_i,Q_j\} \eqa 2 \delta_{ij} H 
-\frac{b}{2} (J_{ik}J_{jk}+J_{jk}J_{ik} -\delta_{ij} J_{kl}J_{kl}) \ .
\label{17-1}
\eea
Note, in particular, that $[Q_i,H]$ vanishes.
This is not a Lie algebra, but rather a quadratically deformed Lie algebra
with $b$ playing the role of deforming parameter.
Of course, as $b$ is proportional to the (A)dS scalar curvature,
in the limit $b\to 0$ one reobtains the flat space constraint algebra.
One may check that this quadratic algebra coincides with the zero mode algebra 
in the Ramond sector of the nonlinear 
$SO(N)$-extended superconformal algebras discovered by 
Bershadsky and Knizhnik in two dimensions 
\cite{Bershadsky:1986ms,Knizhnik:1986wc}.
The above construction gives the quantization of the model obtained at the 
classical level by Kuzenko and Yarevskaya in \cite{Kuzenko:1995mg}.

\section{Geometrical equations for higher spin fields} 
\label{sec4}

We now study the quantum constraints that define the quantization of the 
$O(N)$ spinning particle and use them to derive equations of motion 
for higher spin fields. The case in flat space is well-known, as the
constraints generate the equations of motion of Bargmann and Wigner.
We review this in section \ref{sec:4.1}, though in different language 
and notations, to show how the spinning particle reproduces many of the results 
in higher spin theory, derived previously from field theory.  
More importantly, it indicates how to extend those results to (A)dS 
and conformally flat spaces.
We discuss the extension to (A)dS spaces in section \ref{sec:4.2}. 
For the sake of concreteness, we consider only the case of even $N=2s$,
i.e. massless particles of integer spin $s$.

\subsection{Minkowski space} 
\label{sec:4.1}

In flat space the equations that select the physical states from the 
Hilbert space are given by $T_A |R\ra=0$, where $T_A =(H, Q_i, J_{ij})$ are 
the constraints in (\ref{cons-flatsp}) and $|R\ra$ is
a physical state. We consider even $N=2s$, so that
the constraints can be analyzed by taking complex combinations 
(in a Lorentz invariant way) of the
operators $\psi_i^\mu$, and representing half of them as (Grassmann) coordinates 
and the other half as momenta. Then, one can represent the wave function 
$|R\ra$ in a coordinate basis and expand it in terms of tensors of flat space.
The only tensor surviving the constraints lives in even dimensions $D=2d$, 
has ``$s$'' blocks of ``$d$'' indices
\be
R_{\mu^1_1..\mu^1_d, ..., \mu^s_1..\mu^s_d}
\label{a}
\ee
and satisfies the following three sets of properties:\\
$(i)$ it is symmetric under exchanges of the $s$ blocks,
antisymmetric in the $d$ indices of each block, traceless, and satisfies 
the algebraic Bianchi identities ($J$ constraints);
this part is summarized by saying that the tensor $R$ is an irreducible 
representation of the Lorentz group specified by the Young tableaux 
with $d$ rows and $s$ columns
\bea
&& R_{\mu^1_1..\mu^1_d, ..., \mu^s_1..\mu^s_d} 
\sim
d \quad \underbrace{ \hspace{-11pt}\left \{ \begin{array}{c}\\[-4.5mm]
\yng(4,4,4) 
\end{array}   \right . }_{ s } \quad \ {\rm of}\ \ SO(D-1,1)
\eea
$(ii)$ it satisfies ``differential Bianchi identities'' 
(from half of the $Q$ constraints)
\be
\partial_{[\mu} R_{\mu^1_1..\mu^1_d], ..., \mu^s_1..\mu^s_d} =0 \ ,
\label{b}
\ee
$(iii)$ it satisfies ``Maxwell equations'' (from the other half of the $Q$ constraints)
\be
\partial^{\mu^1_1} R_{\mu^1_1..\mu^1_d, ..., \mu^s_1..\mu^s_d} =0  \ .
\label{c}
\ee
The $H$ constraint is automatically satisfied.
These are geometrical equations for conformal free fields of integer spin $s$,
and are equivalent to the Bargmann-Wigner equations when $D=4$ \cite{Bargmann:1948ck}.
Up to an overall power of the D'Alembertian operator they coincide with the
geometrical equations introduced in \cite{Francia:2002aa1}, that can also be recovered
from the compensator extension of Fronsdal's equations of 
\cite{Francia:2002aa2}.

To derive these equations in more detail, we take complex combinations of 
the $SO(N)=SO(2s)$ indices and define (for $I,i=1,..,s$)
\bea
\psi_I \eqa \frac{1}{\sqrt{2}}(\psi_i + i \psi_{i+s})
\\
\bar \psi_{\bar I} \eqa \frac{1}{\sqrt{2}}(\psi_i - i \psi_{i+s})
\equiv \bar \psi^I
\eea
so that
\be
\{\psi_I^\mu,\bar \psi^{J\nu} \} = \eta^{\mu\nu} \delta_I^J \ .
\ee
In the ``coordinate'' representation one can realize 
$\psi_I^\mu$ as multiplication by Grassmann variables and 
$\bar \psi^I_\mu = {\partial \over \partial \psi_I^\mu }$
(we use left derivatives).
This realization keeps manifest only the $U(s)\subset SO(2s)$ 
subgroup of the internal symmetry group, but will be quite useful 
in classifying the constraints and their solutions.

The susy charges in the $U(s)$ basis take the form
$Q_I= \psi_I^\mu p_\mu$ and $\bar Q^I= \bar \psi^{I \mu} p_\mu$,
and the susy algebra (\ref{3.3}) breaks up into
\be
\{Q_I,\bar Q^J\} = 2 \delta_I^J H \ , 
\qquad \{Q_I,Q_J\}  = \{\bar Q^I,\bar Q^J\} =0 \ .
\label{4.7}
\ee
Similarly, the $SO(N)$ generators 
split as $J_{ij}\sim (J_{I \bar J},J_{IJ}, \bar J_{\bar I \bar J})
\sim (J_I{}^J,K_{IJ}, \bar K^{IJ})$, which we normalize as 
\be
J_I{}^J =\psi_I \cdot \bar \psi^J -d\, \delta_I^J 
\ , \quad
K_{IJ} = \psi_I\cdot \psi_J 
\ , \quad
\bar K^{IJ} = \bar \psi^I \cdot \bar \psi^J  \ ,
\ee
so that $J_I{}^J $ for $I=J$ is a hermitian operator with real eigenvalues.
The $SO(N)$ algebra (\ref{3.5}) breaks up into
\bea
[J_I{}^J,J_K{}^L]\eqa  \delta_K^J J_I{}^L -\delta_I^L  J_K{}^J
\ccr[1mm] 
[J_I{}^J,K_{KL}] \eqa \delta_K^J K_{IL} + \delta_L^J K_{KI}
\ccr[1mm] 
[J_I{}^J,\bar K^{KL}] \eqa - \delta_I^K \bar K^{JL} - \delta_I^L \bar K^{KJ}
\ccr[1mm] 
[K_{IJ},\bar K^{KL}] \eqa  \delta_J^K J_I{}^L - \delta_J^L  J_I{}^K
-\delta_I^K J_J{}^L + \delta_I^L  J_J{}^K
\eea
where the first line identifies the $U(s)$ subalgebra.
Finally, it is useful to list in the same basis the remaining part
of the constraint algebra corresponding to eq. (\ref{3.4})
\bea
[J_I{}^J,Q_K]\eqa \delta_K^J Q_I 
\ccr[1mm] 
[J_I{}^J,\bar Q^K]\eqa - \delta_I^K \bar Q^J 
\ccr[1mm] 
[\bar K^{IJ},Q_K]\eqa  \delta_K^J \bar Q^I  - \delta_K^I \bar Q^J 
\ccr[1mm] 
[K_{IJ},\bar Q^K]\eqa  \delta_J^K Q_I  - \delta_I^K Q_J  \ .
\eea

Let us now analyze the constraint equations, and
derive the geometrical equations for
fields of integer spin $s$, briefly summarized above.
A general wave function is a function of the coordinates $(x^\mu, \psi_I^\mu)$
with a finite Taylor expansion in the Grassmann variables  $\psi_I^\mu$
(with a slight abuse of notation we indicate with $\psi_I^\mu$ both the 
operator and its eigenvalues, but it will be clear from the context 
which is which)
\bea
| R\ra \sim
\sum_{A_i=0}^D
R_{\mu_1..\mu_{A_1}, ...,\, \nu_1..\nu_{A_s}}(x)\, 
\psi_1^{\mu_1}..\psi_1^{\mu_{A_1}}... \psi_s^{\nu_1}..\psi_s^{\nu_{A_s}} \ .
\eea
We start by analyzing the consequences of
the constraints $J_{ij} \sim (J_I{}^J,K_{IJ}, \bar K^{IJ})$.
In the coordinate representation these operators take the form
\be
J_I{}^J =\psi_I \cdot \frac{\partial}{\partial \psi_J} -d\, \delta_I^J 
\ , \quad
K_{IJ} = \psi_I\cdot \psi_J 
\ , \quad
\bar K^{IJ} = \frac{\partial}{\partial \psi_I} \cdot 
\frac{\partial}{\partial \psi_J}
\ee
and we find
\bea  
J_I{}^I |R\ra =0 \ \  ( I \ {\rm fixed}) 
\quad &\Rightarrow& \quad 
|R\ra \sim  R_{\mu_1..\mu_d, ...,\, \nu_1..\nu_d}(x)\, 
\psi_1^{\mu_1}..\psi_1^{\mu_d}... \psi_s^{\nu_1}..\psi_s^{\nu_d} \qquad
\label{8}
\\[1.5mm]
J_I{}^J |R\ra=0  \ \ (I\neq J) 
\quad &\Rightarrow& \quad 
R\ {\rm satisfies\ algebraic\ Bianchi\ identities}
\label{9}
\\[1.5mm]
\bar K^{I J} |R\ra =0
\quad &\Rightarrow& \quad
R\ {\rm traceless}
\label{10}
\\[1.5mm]
K_{IJ} |R\ra =0
\quad &\Rightarrow& \quad 
R\ {\rm traceless\ (in\ dual\ basis)} \ .
\label{11}
\eea
Similarly, the constraints $Q_i=(Q_I, \bar Q^I)$ produce 
\bea
Q_I |R\ra =0  
\quad &\Rightarrow& \quad
 R\ {\rm closed\ (Bianchi\ identities)}
\label{12}
\\[1.5mm]
\bar Q^I |R\ra =0  
\quad &\Rightarrow& \quad 
R\ {\rm co\!-\!closed\ 
(Maxwell\ equations)} \ .
\label{13}
\eea
The constraint $H$ is automatically satisfied as a consequence
of $\{Q_I,\bar Q^J\} =  2 \delta_I^J H $.

Let us comment in more depth some of these equations. 
The constraints (\ref{8}) and  (\ref{9}) correspond to the 
generators of the subgroup 
$U(s) \subset SO(2s)$, which is manifestly realized in  the complex basis.
The curvature $R$ that solves these constraints has ``$s$''
symmetric blocks of ``$d$'' antisymmetric indices each,
and satisfies the algebraic Bianchi identities
\be
R_{[\mu_1..\mu_d,\nu_1]..\nu_d,...} =0   
\ee
where $[...]$ indicates antisymmetrization.
Antisymmetry in each block is manifest.
Symmetry between blocks can be proved by using finite $SO(s) \subset U(s)$ 
rotations. For example, consider the rotation
that exchanges $\psi_I \to \psi_J$ and 
$\psi_J \to -\psi_I$ for fixed $I$ and $J$. This proves symmetry
under exchange of the block relative to the fermions 
$\psi_I$ with the block relative to the fermion $\psi_J$.
As these transformations are connected to the identity,
they are obtained by exponentiating the infinitesimal generators used 
in (\ref{9}), so that this symmetry must be a consequence of (\ref{9}), 
i.e. of the algebraic Bianchi identities. As an aside, we
note that the fermionic Fock vacuum $|\Omega\ra  \sim  \Omega(x)$
is not invariant under the subgroup $[U(1)]^s \subset U(s)$, as the generator
$J_I{}^I$ at fixed $I$ transforms it by an infinitesimal phase
($J_I{}^I |\Omega\ra  = d|\Omega\ra$).
It is the vector $|R\ra$ of eq. (\ref{8}) that is left invariant.
Thus, the constraint $J_I{}^J$ selects an irreducible representation of the
general linear group $GL(D)$ depicted by a Young tableaux
with $d$ rows and $s$ columns. Note that traces are not removed at this stage.

The constraint $\bar K^{IJ}$ removes all possible traces from this tensor, and thus
reduces it to an irreducible representation of the Lorentz group $SO(D-1,1)$.
One may notice that (\ref{11}) (which removes the traces in the dual tensor)
is not independent from (\ref{10}). 
This does not seem to be a consequence of the algebra, but it can be viewed
as a consequence of a duality symmetry enjoyed by the spinning particle.
One can realize the Hodge operator $\star_I$ which takes the dual 
in the $I$-th block of indices by the operation
\be 
\star_I : \psi_I \leftrightarrow \bar \psi^I \ , \qquad (\star_I)^2 = 1   \ .
\ee
This operation can be obtained by a discrete $O(N)$ symmetry transformation
(a reflection on one real $\psi_i$ coordinate).
Denote by ${\star}_{IJ}= {\star}_I {\star}_J $ (this combined 
transformation can be done within $SO(N)$).
Then 
\be
K_{IJ}|R\ra =0 
\quad \Rightarrow \quad 
({\star}_{IJ}\,
K_{IJ}\,
{\star}_{IJ})\,
( {\star}_{IJ}\, |R\ra)
 = \bar K^{I J}  |R^{(\star_{IJ})} \ra  = 0 \ ,
\ee 
which implies that $R^{(\star_{I J})}$
is traceless when contracting an index of the block $I$ with an 
index of the block $J$.  Of course, by $R^{(\star_{I J})}$ we indicate the tensor 
dual to $R$ both in the set of indices of the block $I$ and of the block $J$.
Then, using $\epsilon \epsilon \sim \delta ... \delta$ implies tracelessness 
of $R$ as well.
More generally, invariance under duality implies selfduality, which is 
an expected characterization of conformal field equations in 
higher dimensions, that are precisely those produced by the 
$O(N)$ spinning particle.
Finally, note that (\ref{13}) is a consequence of 
(\ref{12}) and (\ref{10}) (since $ [\bar K^{IJ},Q_K]=
\delta_K^J \bar Q^I - \delta_K^I \bar Q^J$).

\subsubsection{Gauge potentials}

The previous equations can be partially solved and cast in terms of 
gauge potentials for higher spin fields. 
An independent set of constraints that describe the 
geometrical equations is given by (\ref{12}),
(\ref{8})--(\ref{9}), and (\ref{10}), 
corresponding to the constraints $ Q_I, J_I{}^J, \bar K^{IJ}$, 
respectively, and we can try to solve them precisely in that order.

Before starting, it is useful to define the operator 
\be
q = Q_1 Q_2..Q_s
\ee
that satisfies
$ Q_I q = q\, Q_I =0 $ for any $I$. 
In fact, powers of the $Q_I$'s may be nonvanishing up to the $s$-th power,
since an additional application of any of the $Q_I$'s makes it vanish
as a consequence of the algebra (\ref{4.7}).

Constraint (\ref{12}) (i.e. $Q_I|R\ra =0$) 
can be solved by setting
\be
|R\ra = q |\phi\ra \ .
\label{solution-1}
\ee

Constraints (\ref{8})--(\ref{9}) (i.e. $J_I{}^J |R\ra =0$) 
are solved by selecting a tensor 
$R_{\mu^1_1..\mu^1_d, ..., \mu^s_1..\mu^s_d}$
with the symmetries described previously, but not traceless. 
It corresponds to a tensor of $GL(D)$ with a Young tableaux of the form
\bea
R \sim  d \quad \underbrace{ \hspace{-11pt}\left \{ \begin{array}{c}\\[-4.5mm]
 \yng(4,4,4) 
\end{array}   \right . }_{ s } 
\eea
To keep (\ref{8})--(\ref{9}) satisfied by 
(\ref{solution-1}),  one imposes the vanishing of
\be
J_I{}^J q |\phi\ra =  
([J_I{}^J ,q]+ q J_I{}^J )|\phi\ra =  
q (\delta_I{}^J + J_I{}^J  )|\phi\ra = 0
\ee
that is implemented by setting
\be
J_I{}^J |\phi\ra =-  \delta_I{}^J|\phi\ra 
\label{4.26}
\ee
which says that $|\phi\ra$ must have the form
\be
|\phi \ra \sim  \phi_{\mu_1..\mu_{d-1}, ...,\, \nu_1..\nu_{d-1}}(x)\,
\psi_1^{\mu_1}..\psi_1^{\mu_{d-1}}... \psi_s^{\nu_1}..\psi_s^{\nu_{d-1}} 
\ee
and must satisfy corresponding algebraic Bianchi identities. In particular,
the tensor $\phi$ is symmetric under block exchanges.
In short, it corresponds to a Young tableaux of $GL(D)$ of the form
\bea
\phi \sim  d-1\quad 
\underbrace{ \hspace{-11pt}\left \{ \begin{array}{c}\\[-4.5mm]
\yng(4,4) 
\end{array}   \right . }_s  
\label{yt2}
\eea

It remains to implement (\ref{10}) (i.e. $\bar K^{IJ} |R\ra =0$). 
To do this, let us consider
\bea
\bar K^{12}\, q |\phi\ra\eqa
\bar K^{12 }\, 
Q_1 Q_2 Q_3... Q_s
|\phi\ra=   \underbrace{Q_3... Q_s}_{q^{12}}
\bar K^{12}\, 
Q_1 Q_2 |\phi\ra
\ccr
\eqa
q^{12}
\Big [ [\bar K^{12 },  Q_1] Q_2 +
Q_1 [\bar K^{12}, Q_2] + 
Q_1 Q_2 \bar K^{12} \Big]  
|\phi\ra
\ccr
\eqa
q^{12}
\Big [- \bar Q^{2} Q_2 + Q_1  \bar Q^{1} 
+ Q_1 Q_2 \bar K^{12} \Big]  
|\phi\ra
\ccr
\eqa
q^{12}
\Big [ -2H + Q_2 \bar Q^2 + Q_1  \bar Q^1
+ Q_1 Q_2 \bar K^{12} \Big]  
|\phi\ra
\ccr
\eqa
q^{12}
\Big [ -2 H + Q_I  \bar Q^{I} 
+ \frac{1}{2} Q_I Q_J \bar K^{IJ} \Big]  
|\phi\ra
\ccr
\eqa
q^{12} G |\phi\ra
\eea
where we have defined the Fronsdal-Labastida operator\footnote{It corresponds to the
Fronsdal kinetic operator for higher spin fields in $D=4$ \cite{Fronsdal:1978rb}, 
extended to higher dimensions for generic tensors of mixed symmetry 
by Labastida \cite{Labastida:1986ft}.}
\be
G=-2 H + Q_I  \bar Q^{I} + \frac{1}{2}  Q_I Q_J \bar K^{IJ} 
\ee 
which is manifestly $U(s)$ invariant 
(one may check that $[J_I{}^J,G]=0$).
A similar expression holds for $\bar K^{12 }  \to \bar K^{IJ}$,
so that imposing (\ref{10}) produces  (in an obvious notation) 
\be
q^{IJ}\,  G |\phi\ra =0 \ .
\label{eqfl}
\ee
It is convenient to eliminate the operator $q^{IJ}$ form this equation.
Recalling that the product of $s+1$ $Q_I$'s must vanish, one 
finds the following general solution 
\be
G |\phi\ra =  Q_I Q_J Q_K \bar W^K \bar W^J \bar W^I |\rho\ra
\label{24}
\ee
which depends on an arbitrary vector field contained in  
$\bar W^I \equiv W^\mu \bar \psi_\mu^I$, 
and on $|\rho\ra$ that satisfies
$J_I{}^J |\rho\ra =- \delta_I^J|\rho\ra$
(so that it belongs to the same space of 
$|\phi\ra$ and $|\xi\ra$, i.e. it has the same Young tableaux
appearing in eq. (\ref{yt2})).
Eq. (\ref{24}) gives the equations of motion for higher spin fields, 
written in the form that makes use of
the compensator fields described by
$|\rho^{IJK}\ra \equiv\bar W^K \bar W^J \bar W^I |\rho\ra$, 
see \cite{Francia:2002aa2,Sagnotti:2003qa,Bekaert:2002dt,Bandos:2005mb}.

To familiarize with the meaning of the present notation, note
that the effect of $\bar W^I$  acting on $|\rho\ra$
is to saturate one index belonging to the block $I$ 
of the tensor sitting in $|\rho\ra$ 
with the vector field $W^\mu$,
so that $|\rho^{IJK}\ra$ contains a tensor with $s-3$ blocks with 
$d-1$ indices, and the remaining 3 blocks (block $I$, block $J$,  block $K$)
with $d-2$ indices, so that it correspond to a Young tableaux of $GL(D)$ 
of the form
\bea
\rho^{IJK} \sim  d-1\quad 
\underbrace{ \hspace{-11pt}\left \{ \begin{array}{c}\\[-4.5mm]
 \yng(4,1) 
\end{array}   \right . }_s  
\label{yt3}
\eea

Let us now discuss gauge symmetries in this language.
Using an arbitrary vector field $V^\mu(x)$ we define
\be
\bar V^I \equiv V^\mu \bar \psi_\mu^I
\ee
and use it to define the gauge transformation
\be
\delta |\phi\ra = Q_K \bar V^K |\xi\ra \ .
\label{23}
\ee
It is a gauge symmetry of $|R\ra = q |\phi\ra$,
the solution of the Bianchi identities that  
expresses the curvature in terms of the gauge potentials.
Since $[J_I{}^J,Q_K \bar V^K ]=0$, one requires 
that the gauge parameters satisfy 
$J_I{}^J |\xi\ra =- \delta_I^J|\xi\ra$ to guarantee
that  $|\phi\ra$ and $\delta |\phi\ra$ 
are tensors with the same Young tableaux.

To study how the gauge symmetries act on equation (\ref{24}), 
one may compute  the gauge variation of $G |\phi\ra$ using (\ref{23})
\be
G \delta |\phi\ra =-\frac{1}{2} Q_I Q_J Q_K \bar V^K \bar K^{JI} |\xi\ra \ .
\ee
Thus, defining the gauge transformation on the 
compensators as follows
\be
\delta (\bar W^K \bar W^J \bar W^I |\rho\ra) =
-\frac{1}{2} \bar V^{[K} \bar K^{JI]} |\xi\ra
\ee
guarantees gauge invariance of eq. (\ref{24}).

One can use part of the gauge symmetry to set to zero the
compensator fields 
described by $\bar W^K \bar W^J \bar W^I |\rho\ra $,
and obtain the equation of motion in the 
Fronsdal-Labastida  form
\be
G |\phi\ra = 0  \ .
\label{fl-op}
\ee
Inspection of eq. (\ref{24}) indicates that the gauge symmetries surviving 
this partial gauge fixing are those with traceless gauge parameters
$|\xi\ra$, i.e. $\bar K^{IJ} |\xi\ra=0$,
as $\bar K^{IJ}$ in the operator that computes the trace.
For consistency, the gauge potential $|\phi\ra $ must be double traceless.
This can be seen by applying the operator  
$ \bar Q^I -\frac{1}{2} Q_J \bar K^{JI} $
on eq. (\ref{fl-op})
\be 
\Big ( \bar Q^I -\frac{1}{2} Q_J \bar K^{JI} \Big )G |\phi\ra =
-\frac{1}{4} Q_J Q_M Q_N  \bar K^{IJ} \bar K^{MN} |\phi\ra =0
\ee
which is consistent only if $\bar K^{IJ} \bar K^{MN} |\phi\ra =0$,
i.e. if $|\phi\ra$ is double traceless.

In appendix \ref{dictionary}
one finds a dictionary for translating our present notation
to the standard tensorial notation. In particular, one may verify that in $D=4$ 
the gauge potential $|\phi\ra $ corresponds to a symmetric tensor
$\phi_{\mu_1...\mu_s}$, 
the Fronsdal equation 
$ G |\phi\ra \equiv (-2 H + Q_I  \bar Q^{I} + \frac{1}{2}  Q_I Q_J \bar K^{IJ}) |\phi\ra =0$ translates to
\bea
\partial_\alpha \partial^\alpha 
 \phi_{\mu_1...\mu_s} - (\partial_{\mu_1}\partial^{\alpha}
\phi_{\alpha\mu_2...\mu_s} + ...) + (\partial_{\mu_1}\partial_{\mu_2}
\phi^\alpha{}_{\alpha\mu_3...\mu_s} +...) =0 
\eea
where the brackets contain $s$ and $\frac{1}{2} s(s-1)$ terms, respectively,
needed for symmetrizing the $\mu_i$ indices,
and the condition $\bar K^{IJ} \bar K^{MN} |\phi\ra =0$ corresponds 
to $\phi^\alpha{}_\alpha{}^\beta{}_\beta{}_{\mu_5...\mu_s}=0$.

\subsection{(A)dS spaces} 
\label{sec:4.2}

The solutions to the geometrical equations described in the previous
section for Minkowski backgrounds can be deformed to other maximally
symmetric spaces with non-vanishing cosmological constant,
thus producing conformal invariant field equations (see \cite{Metsaev:1995jp}
for an analysis of conformal representations on AdS).
In fact the corresponding constraint algebra, given in eqs. 
(\ref{17}) and (\ref{17-1}), defines a quadratic deformation of
the linear algebra which describes the propagation on flat space, 
and is used to produce the geometrical equations for higher spin fields on (A)dS
spaces. These equations can be worked out, and
correspond to the simple covariantization of the flat space ones,
eqs. (\ref{a}), (\ref{b}), (\ref{c}). They read
\bea
&& R_{\mu^1_1..\mu^1_d, ..., \mu^s_1..\mu^s_d} 
\sim 
d \quad \underbrace{ \hspace{-11pt}\left \{ \begin{array}{c}\\[-4.5mm]
\yng(4,4,4) 
\end{array}   \right . }_{ s }  \quad {\rm of}\ SO(D-1,1)
\ccr 
&&
\nabla_{[\mu} R_{\mu^1_1..\mu^1_d], ..., \mu^s_1..\mu^s_d} =0 
\ccr[1mm]
&&
\nabla^{\mu^1_1} R_{\mu^1_1..\mu^1_d, ..., \mu^s_1..\mu^s_d} =0  
\label{a1} 
\eea
where $\nabla_\mu$ is the covariant derivative on (A)dS spaces.
To analyze them it is again useful to employ a $U(s)$ notation.
The deformed susy algebra reads
\bea
\{ Q_I, Q_J \} &=& b \Big( K_{IL} J_J{}^L +K_{JL} J_I{}^L\Big) 
\label{QI-QJ}
\\[1mm]
\{ \bar Q_I, \bar Q_J \} &=& -b \Big( \bar K_{IL} J_J{}^L +\bar K_{JL}
J_I{}^L\Big) \label{barQI-barQJ} \\[1mm]
\{Q_I,\bar Q^J\} &=& 2 \delta_I^J \Big (H_0- bA_s(D) \Big )
\ccr &&
-\frac{b}{2} \Big ( J_I{}^K J_K{}^J + J_K{}^J J_I{}^K 
-K_{IK}\bar K^{JK} - \bar K^{JK} K_{IK} 
\Big )
\eea
with $A_s(D)=(1-s)\frac{D}{4} -\frac{D^2}{8}$
being the ordering constant given in~(\ref{17}) for the
case $N=2s$, while all other algebraic relations remain unchanged.
Note that in~(\ref{17}) we preferred to use $H$ as hamiltonian to 
make contact with the zero mode sector 
of the Bershadsky-Knizhnik superconformal algebra,
but now we find it more convenient to use  $H_0$, 
which is allowed since the difference is proportional 
to the $J_{ij}$ constraints and the algebra remains first class.
An independent set of constraint is again given by the set 
$ Q_I, J_I{}^J, \bar K^{IJ}$.
We shall discuss in full generality 
the first two constraints, $Q_I$ and  $J_I{}^J$,
which can be solved by the introduction of higher spin gauge potentials.
The main difference with respect to the flat space case is that the 
$Q_I$ operators are no longer anticommuting with one another, so that
$Q_1Q_2\cdots Q_s |\phi\rangle$ 
does not solve the ``Bianchi identity'' constraint anymore 
(the $Q_I$ constraint).

Since $Q_1Q_2\cdots Q_s |\phi\rangle$ 
does solve the Bianchi identity in the flat space limit $b\to 0$, 
we use it as a starting point to 
integrate the higher spin curvature. We find it convenient to use an
explicitly $U(s)$ covariant formulation (actually $SU(s)$ invariant)
and rewrite the above leading order (in powers of $b$) state as 
\bea
|R_0\rangle = q_0 |\phi\rangle\, ,\quad {\rm with}\quad  
q_0\equiv \frac1{s!} \epsilon^{I_1\cdots I_s} Q_{I_1}\cdots Q_{I_s}
\eea 
with the gauge potential $|\phi\rangle\ $  still satisfying
eq. (\ref{4.26}) to solve the $J_I{}^J$ constraint.
Hence, by acting on the previous state with $Q_I$ and by
making repeated use of the anticommutator~(\ref{QI-QJ}), produces on the
right hand side only higher order terms, in powers of $b$. In particular, 
it is not difficult to convince oneself that only operators of the form  
$Q_I\, \epsilon^{I_1\cdots I_s}K_{I_1 I_2}\cdots K_{I_{2n-1} I_{2n}}
Q_{I_{2n+1}}\cdots Q_{I_s}$
are involved. Therefore the higher spin curvature is solved by the
expression 
\bea
|R\rangle = 
\sum_{n=0}^{[s/2]} (-b)^n r_n(s) q_n(s)~|\phi\rangle
\label{SN}
\eea 
where the operators $q_n(s)$ are given by
\bea
q_n(s) \equiv \frac{1}{s!}\epsilon^{I_1I_2\cdots I_s} K_{I_1
  I_2} \cdots K_{I_{2n-1} I_{2n}} Q_{I_{2n+1}}\cdots Q_{I_s}
\eea
and the coefficients $r_n(s)$ are uniquely fixed by imposing the Bianchi identity
(we give a more detailed description of our derivation in the
appendix) and can be written recursively in terms of the Pochhammer function
$P(s,k)\equiv s(s-1)(s-2) \cdots (s-k)$ as follows
\bea
r_n(s) = \frac1{2n}\sum_{k=1}^n r_{n-k}(s)\, a_{2k}(s-2(n-k)+1)       
\ ,  \qquad r_0(s)\equiv 1
\label{rn-1}
\eea  
where
\bea
a_{2k}(s) = f(k) P(s,2k)= f(k)\prod_{l=0}^{2k}(s-l)
\eea
and the $s$-independent function $f(k)$ is defined by
the recursive formula
\bea
f(k)=(-)^k \Bigg [ \frac{1}{(2k+1)!} - \sum_{l=0}^{k-1} \frac{(-)^l}{(2(k-l))!} f(l)\Bigg] 
\ , \qquad f(0)=1 \ .
\label{recf}
\eea
We have checked numerically that these coefficients are generated by
the Taylor expansion of the tangent function, 
$\tan(z)=\sum_{k=0}^\infty f(k) z^{2k+1} $.
This solves the problem of expressing the higher spin curvature in terms of 
gauge potentials on (A)dS spaces.

Note that, alternatively, one may find it more convenient to express
the coefficients~(\ref{rn-1})
in a way that a common Pochhammer function gets factored out, namely
\bea
r_n(s) = \rho_n(s)\ P(s+1,2n)
\eea  
with the prefactor $\rho_n(s)$ given by
\bea
&&\rho_n(s) = \frac{f(n)}{2n}
+\sum_{k_1=1}^{n-1}\frac{f(k_1)f(n-k_1)}{2^2 n(n-k_1)}(s-2n+2k_1+1)\ccr
&&+\sum_{k_1=1}^{n-1}\sum_{k_2=1}^{n-1-k_1}\frac{f(k_1)f(k_2)f(n-k_1-k_2)}{2^3
n(n-k_1)(n-k_1-k_2)} (s-2n+2k_1+1)(s-2n+2k_1+2k_2+1)\ccr&&+\cdots+
\sum_{k_1=1}^{n-1}\sum_{k_2=1}^{n-1-k_1}\cdots \sum_{k_{n-1}=1}^{n-1-k_1\cdots-k_{n-2}}
\frac{f(k_1)f(k_2)\cdots f(n-k_1\cdots-k_{n-1})}{2^n n(n - k_1)\cdots
(n-k_1\cdots-k_{n-1})}\ccr
&&\times (s-2n+2k_1+1)\cdots(s-2n+2k_1\cdots+2k_{n-1}+1)~.
\eea

It remains to study the $\bar K^{IJ}$ constraint, which however
seems rather involved algebraically and we have not attempted to find
a general formula for it.
Nevertheless in the next section we shall treat
explicitly the first few cases, i.e. for spin $s\leq 4$.
Analyses of the geometrical equations for higher spin fields on (A)dS
have been presented also in \cite{Engquist:2007yk,Manvelyan:2007hv}, 
though in the case of totally symmetric potentials that coincide with 
our conformal models only in $D=4$.

Let us conclude this section reporting the explicit expressions 
for the higher spin curvatures for the cases $s\leq 4$. We have
\bea
r_0(s) &=& 1\ccr[2mm]
r_1(s) &=& \frac12 a_2(s+1) = \frac{1}{6}\, (s+1)s(s-1)\ccr[2mm]
r_2(s) &=& \frac14 \biggl( a_4(s+1) +\frac12 a_2(s+1) a_2(s-1)\biggr)
\ccr[2mm] 
&=& \frac{5s +7}{360}\, (s+1)s(s-1)(s-2)(s-3) 
\nonumber
\eea
which provide the following expressions for $s=2,3,4$
\bea
&&|R\rangle = \frac{1}{2!} \epsilon^{I_1I_2} \Big[
Q_{I_1}Q_{I_2} -b\,K_{I_1 I_2}\Big] |\phi\rangle~,
\\[2mm]
&&|R\rangle = \frac{1}{3!} \epsilon^{I_1I_2I_3} \Big[
Q_{I_1}Q_{I_2}Q_{I_3} -4 b\, K_{I_1 I_2} Q_{I_3}\Big] |\phi\rangle~,
\\[2mm]
&&|R\rangle = \frac{1}{4!} \epsilon^{I_1I_2I_3I_4} \Big[
Q_{I_1}Q_{I_2}Q_{I_3}Q_{I_4}-10 b\, K_{I_1 I_2} Q_{I_3}Q_{I_4}+9 b^2
K_{I_1 I_2}K_{I_3 I_4} \Big] |\phi\rangle~.\qquad
\eea

\section{Explicit examples on (A)dS} 

In this section we prove explicitly the gauge invariance on (A)dS backgrounds
of the higher spin curvatures, expressed in terms of gauge potentials, 
for the special cases of spin 2, 3, 4,
and impose the remaining constraints (due to $\bar K^{IJ}$)
that lead to higher derivative equations of motion for the potentials. 
Then we make contact with the standard (quadratic in derivatives)
formulation by introducing compensator fields to maintain the 
gauge invariance of the equations of motion.
Finally we obtain the Fronsdal-Labastida equation for the
double-traceless potentials by gauging to zero the compensators.
\subsection{Spin 2}
The starting point is the $SU(2)$ invariant expression
\bea
|R\rangle  = \frac1{2!} \epsilon^{I_1 I_2} 
\Big [   Q_{I_1}Q_{I_2} -b K_{I_1 I_2}\Big ] |\phi\rangle 
\label{S2}
\eea   
for the spin 2 curvature.
\paragraph{Gauge invariance.}
Let us consider the transformation
\bea\label{2inv}
\delta|\phi\rangle  = Q_K \bar V^K |\xi\rangle 
\eea 
where $\bar V^K = V^a \bar \psi_a^K$ and $|\xi\rangle $ is the
gauge parameter. Both $|\phi\rangle$ and $|\xi\rangle $ 
are described by a rectangular Young tableaux of $GL(D)$
of the type
\bea
\frac{D}{2}-1\quad 
\underbrace{ \hspace{-11pt}\left \{ \begin{array}{c}\\[-4.5mm]
 \yng(2,2,2) 
\end{array}   \right . }_2  
\eea
Now one can easily compute
\be
\delta \biggl(Q_1 Q_2 |\phi\rangle \biggr) = b~ K_{12}~ Q_K \bar
V^K |\phi\rangle 
\ccr[2mm]\qquad \Longrightarrow\qquad \delta |R\rangle  =0~.  
\ee 
This proves that the spin 2 curvature is invariant with respect to 
the gauge transformation (\ref{2inv}).
\paragraph{Equations of motion.}  
The gauge-invariant curvature $|R\rangle $ given above is
expressed in terms of the gauge potential $|\phi\rangle $.
Imposing the left over trace constraint $\bar K^{IJ}
|R\rangle  =0$ produces the equations of motion for the potential. 
We find that
\bea
\bar K^{12}|R\rangle  = G_{2}^{(A)dS}|\phi\rangle =0
\label{EoM:spin2-AdS}
\eea
where we recognize the spin 2 Fronsdal-Labastida kinetic operator on (A)dS
\be
G_{2}^{(A)dS}=\underbrace{-2H_{0}+Q_{I}\bar
Q^{I}+\frac{1}{2}Q_{I}Q_{J}\bar K^{IJ}}_{G}-bK_{IJ}\bar K^{IJ}
+b\alpha_{2}(D) 
\ee
and 
\be
\alpha_{2}(D)=4-\frac
D 2 \left(\frac{D}{2}+1\right)~.
\ee
The operator $G$ looks formally as the one in flat space, but 
of course it is the minimally covariantized version of it.
By expressing the equation of motion~(\ref{EoM:spin2-AdS}) in components 
it is easy to see that, for $D=4$, it reduces to the linearized Einstein 
equation on (A)dS, $R^{(1)}_{\mu\nu}(g+\phi) = 3 b\, \phi_{\mu\nu}$, 
i.e. 
\bea
\nabla^2\phi_{\mu\nu}
-\nabla_{\mu}\nabla^\rho\phi_{\rho\nu}-\nabla_{\nu}\nabla^\rho\phi_{\rho\mu}
+\nabla_\mu\nabla_\nu\phi^\rho{}_\rho
+2b(g_{\mu\nu}\phi^\rho{}_\rho-\phi_{\mu\nu})=0 \ .
\eea
In even dimension $D=2d>4$ it corresponds to
\bea
&&\nabla^2\phi_{\mu_1...\mu_{d-1},\nu_1...\nu_{d-1}}
-(d-1)\Big (\nabla_{\mu_1}\nabla^{\rho}\phi_{\rho\mu_2...\mu_{d-1},\nu_1...\nu_{d-1}}  
+\nabla_{\nu_1}\nabla^{\rho}\phi_{\mu_1\mu_2...\mu_{d-1},\rho\nu_2...\nu_{d-1}}
\Big )
\nonumber\\[1mm] &&
+ (d-1)^2 \nabla_{( \mu_1}\nabla_{\nu_1)} 
\phi^{\rho}{}_{\mu_2...\mu_{d-1},\rho\nu_2...\nu_{d-1}} 
+ 2 b (d-1)^2   g_{\mu_1\nu_1}\phi^{\rho}{}_{\mu_2...\mu_{d-1},\rho\nu_2...\nu_{d-1}}
\nonumber\\[1mm]
&& 
+b \Big (4-d(d+1)\Big ) \phi_{\mu_1...\mu_{d-1},\nu_1...\nu_{d-1}}=0
\eea
where a weighted antisymmetrization in the $\mu$ and $\nu$  groups of indices
is implied and with the round bracket around indices denoting a weighted symmetrization.


\subsection{Spin 3}
We start from the $SU(3)$ invariant expression  
\bea
|R\rangle  = \frac1{3!} \epsilon^{I_1 I_2 I_3} \Big[
  Q_{I_1}Q_{I_2}Q_{I_3} -4b K_{I_1 I_2} Q_{I_3}\Big] |\phi\rangle 
\label{S3}
\eea 
for the spin 3 curvature.
\paragraph{Equations of motion.} Similarly to the spin 2 case we obtain
the equation for the spin 3 potential by imposing
tracelessness of its curvature, $\bar K^{IJ}|R\rangle =0$. 
Using the quadratic algebra described in the previous section, 
we obtain an elegant $U(3)$ covariant result
\bea
&&0=\epsilon_{IKL}\bar K^{KL}|R\rangle  
= Q_I G_{3}^{(A)dS}|\phi\rangle 
\label{EoM:spin3-AdS}
\eea
where 
\be\label{spin3kin}
G_{3}^{(A)dS}=\underbrace{-2H_{0}+Q_{I}\bar
Q^{I}+\frac{1}{2}Q_{I}Q_{J}\bar K^{IJ}}_{G}
-b K_{IJ}\bar K^{IJ} +b\alpha_{3}(D) 
\ee
is the spin 3 Fronsdal-Labastida kinetic operator on (A)dS and 
\be
\alpha_{3}(D)=9-\frac D2 \left(\frac{D}{2}+2\right)~. 
\ee
Note that the equations of motion 
(\ref{EoM:spin3-AdS}) for the spin 3 potential are higher
derivative ones. This is well-known to be correct for 
geometrical equations satisfied by curvatures for spin $s>2$. 
\paragraph{Gauge invariance and Fronsdal-Labastida equation.} 
Using the experience inherited from the flat case, we now study the
gauge invariance and describe the appearance of the compensator field    
$\bar W^{K} \bar W^{J} \bar W^{I}|\rho\rangle $.
First of all, eq. (\ref{EoM:spin3-AdS}) shows that 
$G_{3}^{(A)dS}|\phi\rangle$ is closed with respect the operator
$Q_I$; hence, in analogy with the spin 3 Damour-Deser identity \cite{Damour:1987vm},
one can integrate the $Q_I$  by using the compensator 
to parametrize an element of the kernel of $Q_I$ and 
obtain the searched for second order differential equation 
\be
G_{3}^{(A)dS}|\phi\rangle =\Big(Q_{I}Q_{J}Q_{K}-4b\,
K_{IJ}Q_{K}\Big) \bar W^{K} \bar W^{J} \bar W^{I}|\rho\rangle ~.
\label{EoM:spin3-comp}
\ee
The gauge transformation
\be\label{s3gaugetraf}
\delta|\phi\rangle =Q_{K}\bar{V}^{K}|\xi\rangle 
\ee
is a symmetry of the generalized curvature~(\ref{S3}), whereas the
left hand side of (\ref{EoM:spin3-comp}) transforms as 
\be\label{spin3labtrasf}
G_{3}^{(A)dS}\delta|\phi\rangle =-\frac{1}{2}
(Q_{I}Q_{J}Q_{K}-4b\, K_{IJ}Q_{K})\bar{V}^{[K} \bar K^{JI]}|\xi\rangle ~. 
\ee
Hence, the differential equation with compensator is fully 
gauge-invariant provided the compensator transforms as
\be
\delta (\bar W^{K} \bar W^{J} \bar W^{I}|\rho\rangle )=
-\frac{1}{2}\bar{V}^{[K} \bar K^{JI]}|\xi\rangle ~.
\label{transf:comp}
\ee
The latter can be used at once to gauge fix the
 compensator to zero yielding 
 \be
G_{3}^{(A)dS}|\phi\rangle =0
\ee
that is the second order spin 3 
Fronsdal-Labastida equation on (A)dS.
The left over gauge symmetry must keep the
left hand side of~(\ref{transf:comp}) equal to zero,
$\bar{V}^{[K} \bar K^{JI]}|\xi\rangle=0$.  Hence, the gauge parameter must 
be traceless. 
\subsection{Spin 4}
We start from the manifestly $SU(4)$ invariant expression   
\bea
|R\rangle  = \frac1{4!} \epsilon^{I_1 I_2 I_3 I_4} \Big[
  Q_{I_1}Q_{I_2}Q_{I_3}Q_{I_4} - 10 b ~ K_{I_1 I_2} Q_{I_3} Q_{I_4} +9b^2
   ~ K_{I_1 I_2}K_{I_3 I_4}\Big] |\phi\rangle 
\label{S4}
\eea 
for the spin 4 curvature. 
\paragraph{Equations of motion.} 
The traceless condition (in the form 
$\epsilon_{IJKL}\bar K^{KL}|R\rangle =0$) 
produces the higher order equations of motion 
\be
\Big(Q_{[I}Q_{J]}-bK_{IJ}\Big)G_{4}^{(A)dS}|\phi\rangle =0
\label{EoM:spin4-AdS}
\ee
where
\be
G_{4}^{(A)dS} = 
\underbrace{-2H_{0}+Q_{I}\bar Q^{I}+\frac{1}{2}Q_{I}Q_{J}\bar K^{IJ}}_{G} 
-bK_{IJ}\bar K^{IJ}+b\alpha_{4}(D)
\ee
is the second order Fronsdal-Labastida differential operator on (A)dS and
\be
\alpha_{4}(D)=16-\frac D 2 \left(\frac{D}{2}+3\right)~.
\ee  

\paragraph{Gauge invariance and Fronsdal-Labastida equation.} 
Once again the higher order equations of motion (\ref{EoM:spin4-AdS})
are fully gauge invariant under
$\delta|\phi\rangle =Q_{K}\bar{V}^{K}|\xi\rangle$. On the other hand
it is straightforward to check that, identically to the spin 3 case,
one gets
\be\label{spin4labtrasf}
G_{4}^{(A)dS}\delta|\phi\rangle =-\frac{1}{2}
(Q_{I}Q_{J}Q_{K}-4b\, K_{IJ}Q_{K})\bar{V}^{[K} \bar K^{JI]}|\xi\rangle 
\ee
so that the ``compensated'' second  order equation
\be
G_{4}^{(A)dS}|\phi\rangle =\Big(Q_{I}Q_{J}Q_{K}-4b\,
K_{IJ}Q_{K}\Big) \bar W^{K} \bar W^{J} \bar W^{I}|\rho\rangle
\label{EoM:spin4-comp}
\ee
is invariant, provided the compensator transforms as
in~(\ref{transf:comp}). The Fronsdal-Labastida equation
\be
G_{4}^{(A)dS}|\phi\rangle =0
\ee
is again obtained by gauge fixing the compensator to zero. 
It is invariant under gauge transformations parametrized by a traceless
parameter and requires a gauge potential with vanishing double trace. 

\subsection{Spin $s > 4$ }
The results obtained above suggest us that, for every integer spin $s$ 
in arbitrary (even) dimensions $D$, the Fronsdal-Labastida kinetic operator
on (A)dS becomes
\be\label{fronads}
G_{s}^{(A)dS}=\Biggl[
-2H_{0}+Q_{I}\bar Q^{I}+\frac{1}{2}Q_{I}Q_{J}\bar K^{IJ}
-bK_{IJ}\bar K^{IJ}+b\alpha_{s}(D) \Biggr]
\ee
where
\be
\alpha_{s}(D)=s^{2}-\frac{D}{2}\left(\frac{D}{2}+s-1\right)=s^2+ 2 A_s(D)~.
\ee
One can check that the  gauge transformation of $G_{s}^{(A)dS}|\phi\rangle$ 
is identical to the ones obtained above in (\ref{spin3labtrasf}) 
and (\ref{spin4labtrasf}) for spin 3 and spin 4, respectively, 
and it is gauge invariant provided the gauge parameter is traceless. 
Moreover, in $D=4$ this operator reproduces the extension
of the Fronsdal operator to (A)dS spaces.

\section{Conclusions}

We have discussed classical and quantum properties of the 
$O(N)$ spinning particles and studied their relation to the equations of motion
for fields of spin $s=\frac{N}{2}$.
After a review of the model, we have shown how these spinning particles
can be coupled to conformally flat spaces, both  classically 
and quantum mechanically, thus extending the result of \cite{Kuzenko:1995mg},
where the coupling to (A)dS spaces was obtained at the classical level. 
One of our results, worth mentioning, is that on (A)dS the algebra of 
quantum constraints closes quadratically
and reproduces the zero mode sector of the 2D Bershadsky-Knizhnik $SO(N)$-extended 
nonlinear superconformal algebra \cite{Bershadsky:1986ms,Knizhnik:1986wc}.

Furthermore, we have analyzed the constraint equations that select 
the physical states from the particle Hilbert space.
We have shown that in flat space these equations reproduce
the so-called  geometrical equations for higher spin curvatures.
Using the quantum mechanical operators we have described how to integrate
the ``Bianchi identities'' to express curvatures in term
of gauge potentials, and obtained various well-known forms of the 
equations of motion for higher spin fields 
\cite{Fronsdal:1978rb,Ouvry:1986dv,Siegel:1986zi,Labastida:1986ft,
Francia:2002aa2,Sagnotti:2003qa, Bekaert:2002dt,de Medeiros:2002ge,Bandos:2005mb,
Sorokin:2004ie}.

Then we have studied the spinning particles on (A)dS spaces and obtained 
corresponding geometrical equations. To our knowledge generalized Poincar\'e
lemmas are not known for this case, but using the constraint algebra
we have shown how to integrate the ``Bianchi identities'' in terms of gauge potentials.
Finally, we have analyzed in detail the equations of motion
and the gauge invariances for the cases of spin $s \leq 4$. 

Having established the precise connection between the quantum theory of
the $O(N)$ spinning particles and the conformal higher spin field equations on (A)dS, 
one can now use the equivalent path integral quantization to obtain
further results on the quantum theory of higher spin fields.

\acknowledgments{
FB would like to thank the organizers of the Simons Workshop 
in Mathematics and Physics 2008, Stony Brook University and, in particular, 
Warren Siegel for discussions. 
We also thank Augusto Sagnotti and Dmitri Sorokin for discussions.
This work was supported in part by the Italian MIUR-PRIN contract 20075ATT78.} 
\vskip2cm


\begin{appendix}

\section{Dictionary}
\label{dictionary}
For the reader's convenience, we present a dictionary between 
our compact notation and the more conventional tensorial notation.  
Building blocks are the superalgebra constraints that lead to
the geometrical equations
\bea
Q_I &=& -i\psi^a_I e^{\mu}_a \Big( \partial_{\mu}+\omega_{\mu bc}
\psi^b_J \frac{\partial}{\partial \psi_{Jc}}\Big)\,,\quad\quad
\bar Q^I = -i\frac{\partial}{\partial\psi_{Ia}} 
e^{\mu}_a \Big( \partial_{\mu}+\omega_{\mu bc}
\psi^b_J \frac{\partial}{\partial \psi_{Jc}}\Big)\nonumber\\[2mm]
J_I{}^J
&=& \psi_I^a\frac{\partial}{\partial\psi^a_J}-d\delta_I^J\,,\quad
K_{IJ} =\psi_I^a \psi_{Ja}\,,\quad \bar K^{IJ}
=\frac{\partial}{\partial\psi^a_I}\frac{\partial}{\partial\psi_{Ja}}~. 
\nonumber 
\eea
As an example, let us consider a state corresponding to a rectangular tensor
\bea
|X\rangle =  X_{a_1..a_n,b_1 .. b_n,\ldots, c_1..
 c_n}\psi_1^{a_1}..\psi_1^{a_n}\ \psi_2^{b_1}..\psi_2^{b_n}\
\ldots\ \psi_s^{c_1}..\psi_s^{c_n}
\sim  n\quad 
\underbrace{ \hspace{-11pt}\left \{ \begin{array}{c}\\[-4.5mm]
\yng(4,4) 
\end{array}   \right . }_s  
\nonumber
\eea
with $n$ arbitrary (and similar expansions for more general tensors).
A set of correspondences that allows to obtain
Fronsdal-Labastida equations in components, is given by  
\begin{eqnarray}
\begin{array}{l|l}
\ {\rm Compact\ notation}\ & \ {\rm Tensorial\ notation}
\\\hline\hline 
\ \phantom{\Big(}  |X\rangle\ &\ X_{a_1.. a_n, b_1..b_n,\ldots, c_1.. c_n}
\\[2mm] \hline
\ \phantom{\Big(}  Q_1|X\rangle\ &\ -i\nabla_{a_1}X_{a_2..
a_{n+1}, b_1.. b_n,\ldots, c_1.. c_n}   
\\[2mm] \hline
\ \phantom{\Big(} \bar Q^1|X\rangle\ &\ -in\nabla^{l}X_{l
a_2.. a_{n}, b_1.. b_n,\ldots, c_1.. c_n}   
\\[2mm] \hline
\ \phantom{\Big(} \bar K^{12}|X\rangle\ &\ (-)^nn^2\,
X^l{}_{a_2.. a_{n}, l b_2.. b_n,\ldots, c_1.. c_n}   
\\[2mm] \hline
\ \phantom{\Big(}  K_{12}|X\rangle\ &\
(-)^n\eta_{a_1b_1}X_{a_2..a_{n+1}, b_2..b_{n+1},\ldots,
c_1.. c_n}   
\\[2mm] \hline
\ \phantom{\Big(}  J_1{}^1|X\rangle\ &\ (n-d)X_{a_1.. a_{n}, b_1.. b_n,\ldots, c_1.. c_n}  
\\[2mm] \hline
\end{array}
\nonumber
\end{eqnarray}
so that
\begin{eqnarray}
\begin{array}{l|l}
\ {\rm Compact\ notation}\ & \ {\rm Tensorial\ notation}
\\\hline\hline 
\  \phantom{\Big(}  H_0|X\rangle\  &\  -\frac12 \nabla^2X_{a_1..
  a_n, b_1.. b_n,\ldots, c_1.. c_n} 
\\[2mm] \hline
\  \phantom{\Big(}  Q_I \bar Q^I |X\rangle\  &\  -n\Big(\nabla_{a_1}\nabla^l X_{l 
a_2.. a_n, b_1.. b_n,\ldots, c_1.. c_n} \  \\ & 
\hskip1cm +
\nabla_{b_1}\nabla^l X_{a_1.. a_n, l b_2.. b_n,\ldots,
  c_1 .. c_n} \\ & 
\hskip1cm
+\cdots +\nabla_{c_1}\nabla^l X_{a_1.. a_n, b_1 .. b_n,\ldots,l c_2.. c_n}\Big) 
\\[2mm] \hline
\  \phantom{\Big(}  Q_I Q_J \bar K^{IJ} |X\rangle\  &\ 2n^2\Big(\nabla_{(a_1}\nabla_{b_1)} X^l{}_{ 
a_2.. a_n, l b_2.. b_n,\ldots, c_1.. c_n}\  \\ &
\hskip1cm
+\cdots +
\nabla_{(a_1}\nabla_{c_1)} X^l{}_{a_2.. a_n, b_1.. b_n,\ldots,l
  c_2.. c_n}\  \\ &
\hskip1cm
+\cdots +\nabla_{(b_1} \nabla_{c_1)} X_{a_1.. a_n,}{}^l{}_{b_2.. b_n,\ldots,l c_2.. c_n}+\cdots\Big) 
\\[2mm]\hline
\  \phantom{\Big(}  K_{IJ} \bar K^{IJ} |X\rangle\  &\ -2n^2\Big(\eta_{a_1 b_1} X^l{}_{ 
a_2.. a_n, l b_2.. b_n,\ldots, c_1 .. c_n}\  \\ &\hskip1.2cm +\cdots +
\eta_{a_1 c_1} X^l{}_{a_2.. a_n, b_1.. b_n,\ldots, l c_2.. c_n}\  \\ &
\hskip1.2cm
+\cdots +\eta_{b_1 c_1} X_{a_1.. a_n,}{}^l{}_{b_2.. b_n,\ldots, l c_2.. c_n}+\cdots\Big)\  
\\[2mm]\hline
\end{array}
\nonumber
\end{eqnarray}
where a weighted antisymmetrization in each of the $s$ groups of indices
$a_i,b_i,\cdots,c_i$ is implied. In the last two expressions 
the dots in parenthesis indicate a sum over all pairs of indices corresponding to $I<J$
and the round brackets around indices denote a weighted symmetrization.

\section{Solution to the ``Bianchi identities'' on (A)dS}
\label{App1} 

We give here a detailed derivation of the 
solution to the ``Bianchi identities'' equations for the
higher spin curvatures on (A)dS. In the spinning particle language
such equations read
\bea
&& J_I{}^J |R\rangle =0\\
&& Q_I |R\rangle =0\,,\quad I,J=1,\dots,s~.
\eea
As explained in the main text the first relation select an
irreducible  $GL(D)$ tensor represented 
by a rectangular Young tableaux with $s$ rows and $D/2$ columns. 
The ``differential Bianchi identity'' is instead encoded in the second
relation, and can be solved by expressing the curvature $|R\rangle$  
in terms of a potential $|\phi\rangle$
\bea
|R\rangle = q|\phi\rangle
\eea    
where the operator $q$ must reduce in the flat space limit to 
\bea
q 
\quad \stackrel{\rm flat\ space}{\longrightarrow}\quad Q_1
Q_2 \cdots Q_s =\frac1{s!} \epsilon^{I_1\cdots I_s}
Q_{I_1}\cdots Q_{I_s} \equiv q_0
\eea
and, since $[J_I{}^J, Q_K]= \delta_K^J\, Q_I$, the  potential must satisfy
\bea
J_I{}^J |\phi\rangle=-\delta_I^J |\phi\rangle
\label{alg-bianchi}
\eea
so that is represented by a Young tableaux with $s$ columns
and $D/2 -1$ rows. Above and
in what follows we express the differential operator $q$ in an
explicitly $SU(s)$ invariant form. 
We construct $q$ by imposing the conditions
\bea
Q_I|R\rangle =0
\eea
and use its flat space limit $q_0$ as our starting point.
In particular, thanks to the $SU(s)$-invariance it will
suffice to require $Q_1|R\rangle =0$. In order to achieve such a task we
shall need a few recursive relations that we derive using the
commutation relations  
\bea
\{Q_I, Q_J\} &=& b \left( K_{IL} J_J{}^L +K_{JL} J_I{}^L \right)
\\[2mm]
[J_I{}^J, Q_K]&=& \delta_K^J ~Q_I  
\\[2mm]
[K_{IJ} , K_{KL}] &=& [K_{IJ},Q_K] =0
\eea  
 and the condition~(\ref{alg-bianchi}). We find it convenient to split the $s$
 indices into a ``time-like'' index $1$ and $s-1$ ``space-like''
 indices $i$
\bea
I =(1,i)~,\quad i=2,\dots, s~.
\eea 
Let us define a shortcut notation that will prove to be extremely useful
\bea
&& \epsilon^{i_1\cdots i_{s-1}} ~Q_{i_1}\cdots Q_{i_n} Q_1
Q_{i_{n+1}}\cdots Q_{i_{s-1}}\quad \longrightarrow\quad
Q_{[n]}Q_1 Q_{[s-1-n]}\ccr
&& \epsilon^{i_1\cdots i_{s-1}} ~K_{1 i_1}~ Q_{i_3}\cdots Q_{i_n} Q_1
 Q_{i_{n+1}}\cdots Q_{i_{s-1}}\quad \longrightarrow\quad
K_{1 i_1}~Q_{[n-2]}Q_1 Q_{[s-1-n]}
\nonumber
\eea
and whenever we encounter a $K_{ab}$ tensor we use the commutation
rules above and the antisymmetry provided by the $\epsilon$ tensor 
to bring it in front of everything and give it the first indices 
of the set $i_1, i_2,\dots$.  
It is thus not difficult to prove the relation
\bea
(-)^n Q_{[n]} Q_1 Q_{[s-1-n]}|\phi\rangle &=& Q_1 Q_{[s-1]}|\phi\rangle 
+b\biggl(
  n(s-2)-\frac{n(n-1)}{2}\biggr) K_{1 i_1} Q_{[s-2]}|\phi\rangle \nonumber\\
&&-b K_{i_1 i_2}
\sum_{m=1}^n \sum_{k=m-1}^{s-3} (-)^{k} Q_{[k]} Q_1
Q_{[s-k-3]}|\phi\rangle
\eea
that can be iterated by noting that the last term is just equal
to the left hand side provided one performs the substitution $s\to s-2$. The
iteration process thus yields
\bea
&& \sum_{n=0}^{s-1} (-)^n Q_{[n]} Q_1 Q_{[s-1-n]}|\phi\rangle = s
Q_1 Q_{[s-1]}|\phi\rangle
\ccr
&& -(-b)\, a_2(s) \biggl( K_{1 i_1} Q_{[s-2]} -K_{i_1 i_2} Q_1
Q_{[s-3]}\biggr)|\phi\rangle
\ccr && -(-b)^2 a_4(s) \biggl( K_{1 i_1} K_{i_2 i_3}Q_{[s-4]} -K_{i_1
  i_2} K_{i_3 i_4} Q_1 Q_{[s-5]}\biggr)|\phi\rangle
\ccr && .\ccr && .
\ccr && -(-b)^p a_{2p}(s) K_{1 i_1} K_{i_2 i_3}\cdots K_{i_{2(p-1)}
  i_{2p-1}} Q_{[s-2p]}|\phi\rangle\ccr &&
+(-b)^p \sum_{k_0=1}^{s-1} \sum_{m_1=1}^{k_0} \sum_{k_1=m_1-1}^{s-3}
\cdots \sum_{m_p=1}^{k_{p-1}} \sum_{k_p=m_p-1}^{s-2p-1} (-)^{k_p}\ccr
&&\hskip1cm K_{1 i_1} K_{i_2 i_3}\cdots K_{ i_{2(p-1)}i_{2p-1}} Q_{[k_p]} Q_1
Q_{[s-2p-1-k_p]} |\phi\rangle  
\label{master1}
\eea
with
\bea
a_{2n}(s)&\equiv &\sum_{k_0=1}^{s-1} \sum_{m_1=1}^{k_0} \sum_{k_1=m_1-1}^{s-3}
\cdots \sum_{m_n=1}^{k_{n-1}} \sum_{k_n=m_n-1}^{s-2n-1} 1
= f(n)\, P(s,2n)
\eea
where $P(s,2n) = s(s-1)\cdots (s-2n)$ is the Pochhammer function and
the $s$-independent function $f(n)$ is given be the recursive formula
(equivalent to (\ref{recf}))
\bea
\sum_{k=0}^{n} \frac{(-)^k}{(2k)!}f(n-k)=\frac{(-)^n}{(2n+1)!}~.
\eea
Note that the iterative expression~(\ref{master1}) stops at the
last-but-one entry if $s=2p$, whereas it stops at the last entry if $s=2p+1$. 
Another helpful relation that can be obtained with the help
of~(\ref{master1}) and with implied antisymmetrization of the indices
``$i$'', reads 
\bea
Q_1^2 Q_{[s-1]}|\phi\rangle &=& bK_{1 i_1} \sum_{n=0}^{s-2} (-)^n Q_{[n]} Q_1
Q_{[s-2-n]} |\phi\rangle\ccr
&=& bK_{1 i_1} \biggl(
a_0(s-1)\, Q_1Q_{[s-2]} -b\, a_2(s-1)\,K_{i_2 i_3} Q_1 Q_{[s-4]}  
\ccr &&\hskip1.5cm +\cdots
+(-b)^{p-1} \,a_{2(p-1)}(s-1) \, K_{i_2 i_3}\cdots K_{i_{2(p-1)} i_{2p-1}}
Q_1 \biggr)|\phi\rangle~.\ccr &&
\label{master2}
\eea
It is easy now to convince oneself that the zero-th order operator
$q_0(s)$ can be written as
\bea
s~ q_0 (s) = \sum_{n=0}^{s-1} (-)^n Q_{[n]} Q_1
Q_{[s-1-n]}
\eea
so that making use of~(\ref{master1}), (\ref{master2}), and assuming for 
definiteness that $s=2p$, one gets
\bea
s!~ Q_1 q_0(s) |\phi\rangle &=& -\sum_{n=1}^{s/2}(-b)^n
a_{2n}(s+1) Q_1 I_n(s)|\phi\rangle
\label{1step}
\eea 
where
\bea
I_n(s) \equiv K_{1 i_1} K_{i_2 i_3}\cdots K_{i_{2(n-1)} i_{2n-1}} Q_{[s-2n]} 
\eea
and we have used the identity
\bea
\sum_{k=0}^n a_{2k}(s) a_{2(n-1-k)}(s-1-2k)=a_{2n}(s+1)\,,
\quad\quad a_{-2}(s)\equiv 1
\eea
that can be proved
by induction. This completes the first step. The next step
is to rewrite expression~(\ref{1step}) in terms of
$U(s)$-covariant tensors. The covariantization of the tensors
$I_n(s)$ is again an iterative process. Note in fact that one can
write
\bea
I_n(s)|\phi\rangle = \frac {s!}{2n} q_n(s)|\phi\rangle 
+\frac1{2n}\sum_{m=n+1}^{s/2} 
(-b)^{m-n} ~a_{2(m-n)}(s-2n+1)~ I_m(s)|\phi\rangle
\nonumber
\eea 
that finally yields 
\bea
Q_1 ~ \sum_{n=0}^{[s/2]}(-b)^n r_n(s) q_n(s) |\phi\rangle =0  
\label{2step}
\eea
with
\bea
r_n(s) = \frac1{2n}\sum_{k=1}^n r_{n-k}(s)~a_{2k} (s-2(n-k)+1) ~.
\eea
Finally, note that in~(\ref{2step}) we have replaced $s/2$ with its integer
part: it is in fact not difficult to check that the latter holds for
odd $s$ as well, with that precise replacement.

\end{appendix} 

\vskip1cm



\begin{thebibliography}{99} 
 
\bibitem{Bastianelli:2007pv}
  F.~Bastianelli, O.~Corradini and E.~Latini,
  JHEP {\bf 0702} (2007) 072
  [arXiv:hep-th/0701055].

\bibitem{Gershun:1979fb} 
V.~D.~Gershun and V.~I.~Tkach, 
Pisma Zh.\ Eksp.\ Teor.\ Fiz.\  {\bf 29} (1979) 320 
[Sov. Phys. JETP {\bf 29} (1979) 288]. 

\bibitem{Howe:1988ft} 
P.~S.~Howe, S.~Penati, M.~Pernici and P.~K.~Townsend, 
Phys.\ Lett.\ B {\bf 215} (1988) 555.

\bibitem{Howe:1989vn} 
P.~S.~Howe, S.~Penati, M.~Pernici and P.~K.~Townsend, 
Class.\ Quant.\ Grav.\  {\bf 6} (1989) 1125. 

\bibitem{Sorokin:2004ie}
M. A. Vasiliev, Higher Spin Gauge Theories in Various Dimensions,
Fortsch. Phys. {\bf 52} (2004) 702,  [arXiv:hep-th/0401177];\\
D. Sorokin, Introduction to the Classical Theory of Higher Spins, AIP
Conf. Proc. {\bf 767} (2005) 172 (2005), [arXiv:hep-th/0405069];\\
N. Bouatta, G. Compere and A. Sagnotti, An Introduction to Free
Higher-Spin Fields, [arXiv:hep-th/0409068];\\
X. Bekaert, S. Cnockaert, C. Iazeolla and M. A. Vasiliev, Nonlinear
higher spin theories in various dimensions, [arXiv:hep-th/0503128];\\
A.~Fotopoulos and M.~Tsulaia,
Gauge Invariant Lagrangians for Free and Interacting Higher Spin
Fields. A Review of the BRST formulation.
 arXiv:0805.1346 [hep-th].

\bibitem{Porrati:2008rm}
  M.~Porrati,
  arXiv:0804.4672 [hep-th].

\bibitem{Benincasa:2007xk}
  P.~Benincasa and F.~Cachazo,
  arXiv:0705.4305 [hep-th].

\bibitem{Kuzenko:1995mg} 
  S.~M.~Kuzenko and Z.~V.~Yarevskaya, 
  Mod.\ Phys.\ Lett.\ A {\bf 11} (1996) 1653 
  [arXiv:hep-th/9512115]. 

\bibitem{Siegel:1988ru} 
  W.~Siegel, 
  Int.\ J.\ Mod.\ Phys.\ A {\bf 3}, 2713 (1988);
  Int.\ J.\ Mod.\ Phys.\ A {\bf 4}, 2015 (1989). 

\bibitem{Marnelius:1978fs}
  R.~Marnelius,
  Phys.\ Rev.\  D {\bf 20} (1979) 2091.

\bibitem{Bastianelli:2006rx}
F.~Bastianelli and P.~van Nieuwenhuizen,
{\em ``Path integrals and anomalies in curved space,''}
(Cambdridge University Press, 2006).

\bibitem{Bastianelli:2005rc}
 F.~Bastianelli,
{\em  ``Path integrals in curved space and the worldline formalism,''}
proceedings of ``Path integrals: from quantum information to cosmology'' 
(Prague, 2005), arXiv:hep-th/0508205.

\bibitem{Bastianelli:2002fv}
  F.~Bastianelli and A.~Zirotti,
  Nucl.\ Phys.\  B {\bf 642} (2002) 372
  [arXiv:hep-th/0205182].

\bibitem{Bastianelli:2002qw}
  F.~Bastianelli, O.~Corradini and A.~Zirotti,
  Phys.\ Rev.\  D {\bf 67} (2003) 104009
  [arXiv:hep-th/0211134];
  JHEP {\bf 0401} (2004) 023
  [arXiv:hep-th/0312064].

\bibitem{Bastianelli:2005vk}
  F.~Bastianelli, P.~Benincasa and S.~Giombi,
  JHEP {\bf 0504} (2005) 010
  [arXiv:hep-th/0503155];
  JHEP {\bf 0510} (2005) 114
  [arXiv:hep-th/0510010].

\bibitem{Francia:2002aa1}
  D.~Francia and A.~Sagnotti,
  Phys.\ Lett.\  B {\bf 543} (2002) 303
  [arXiv:hep-th/0207002].

\bibitem{Francia:2002aa2}
D.~Francia and A.~Sagnotti,
  Class.\ Quant.\ Grav.\  {\bf 20} (2003) S473
  [arXiv:hep-th/0212185];
  Phys.\ Lett.\  B {\bf 624} (2005) 93
  [arXiv:hep-th/0507144].

\bibitem{Weinberg:1965rz}
  S.~Weinberg,
  Phys.\ Rev.\  {\bf 138} (1965) B988.

\bibitem{de Wit:1979pe}
  B.~de Wit and D.~Z.~Freedman,
  Phys.\ Rev.\  D {\bf 21} (1980) 358.

\bibitem{Sagnotti:2003qa}
  A.~Sagnotti and M.~Tsulaia,
  Nucl.\ Phys.\  B {\bf 682} (2004) 83
  [arXiv:hep-th/0311257];
  D.~Francia, J.~Mourad and A.~Sagnotti,
  Nucl.\ Phys.\  B {\bf 773} (2007) 203
  [arXiv:hep-th/0701163].

\bibitem{DuboisViolette:2001jk}
  M.~Dubois-Violette and M.~Henneaux,
Lett. Math. Phys. {\bf 49} (1999) 245
[arXiv:math/9907135];
  Commun.\ Math.\ Phys.\  {\bf 226} (2002) 393
  [arXiv:math/0110088].

\bibitem{Bekaert:2002dt}
  X.~Bekaert and N.~Boulanger,
  Commun.\ Math.\ Phys.\  {\bf 245} (2004) 27
  [arXiv:hep-th/0208058];
  Phys.\ Lett.\  B {\bf 561} (2003) 183
  [arXiv:hep-th/0301243].

\bibitem{de Medeiros:2002ge}
  P.~de Medeiros and C.~Hull,
  Commun.\ Math.\ Phys.\  {\bf 235} (2003) 255
  [arXiv:hep-th/0208155];
  JHEP {\bf 0305} (2003) 019
  [arXiv:hep-th/0303036].

\bibitem{Bandos:2005mb}
  I.~Bandos, X.~Bekaert, J.~A.~de Azcarraga, D.~Sorokin and M.~Tsulaia,
  JHEP {\bf 0505} (2005) 031
  [arXiv:hep-th/0501113].

\bibitem{Fronsdal:1978rb}
  C.~Fronsdal,
  Phys.\ Rev.\  D {\bf 18} (1978) 3624.

\bibitem{Labastida:1986ft}
  J.~M.~F.~Labastida,
  Phys.\ Rev.\ Lett.\  {\bf 58} (1987) 531;
  Nucl.\ Phys.\  B {\bf 322} (1989) 185.

\bibitem{Bershadsky:1986ms}
  M.~A.~Bershadsky,
  Phys.\ Lett.\  B {\bf 174} (1986) 285.

\bibitem{Knizhnik:1986wc}
  V.~G.~Knizhnik,
  Theor.\ Math.\ Phys.\  {\bf 66} (1986) 68
  [Teor.\ Mat.\ Fiz.\  {\bf 66} (1986) 102].

\bibitem{Buchbinder:2001bs}
  I.~L.~Buchbinder, A.~Pashnev and M.~Tsulaia,
  Phys.\ Lett.\  B {\bf 523} (2001) 338
  [arXiv:hep-th/0109067].

\bibitem{Bandos:1998vz}
  I.~A.~Bandos and J.~Lukierski,
  Mod.\ Phys.\ Lett.\  A {\bf 14} (1999) 1257
  [arXiv:hep-th/9811022].

\bibitem{Bandos:1999qf}
  I.~A.~Bandos, J.~Lukierski and D.~P.~Sorokin,
  Phys.\ Rev.\  D {\bf 61} (2000) 045002
  [arXiv:hep-th/9904109].

\bibitem{Hallowell:2007qk}
  K.~Hallowell and A.~Waldron,
  Commun.\ Math.\ Phys.\  {\bf 278} (2008) 775 
  [arXiv:hep-th/0702033];
SIGMA 3 (2007) 089  [arXiv:0707.3164 [math.DG]].


\bibitem{Pashnev:1997rm}
  A.~Pashnev and M.~M.~Tsulaia,
  Mod.\ Phys.\ Lett.\  A {\bf 12} (1997) 861
  [arXiv:hep-th/9703010],
  Mod.\ Phys.\ Lett.\  A {\bf 13} (1998) 1853
  [arXiv:hep-th/9803207].

\bibitem{Henneaux:1985kr}
  M.~Henneaux,
  Phys.\ Rept.\  {\bf 126} (1985) 1.

\bibitem{Bargmann:1948ck}
  V.~Bargmann and E.~P.~Wigner,
  Proc.\ Nat.\ Acad.\ Sci.\  {\bf 34} (1948) 211.

\bibitem{Marnelius:1988ab}
  R.~Marnelius and U.~Martensson,
  Nucl.\ Phys.\  B {\bf 321} (1989) 185.

\bibitem{Siegel:1999ew}
  W.~Siegel,
  ``Fields,'' chapter XII,
  arXiv:hep-th/9912205.


\bibitem{Metsaev:1995jp}
  R.~R.~Metsaev,
  Mod.\ Phys.\ Lett.\  A {\bf 10} (1995) 1719.

\bibitem{Engquist:2007yk}
  J.~Engquist and O.~Hohm,
  JHEP {\bf 0804} (2008) 101
  [arXiv:0708.1391 [hep-th]].

\bibitem{Manvelyan:2007hv}
  R.~Manvelyan and W.~Ruhl,
  Nucl.\ Phys.\  B {\bf 797} (2008) 371
  [arXiv:0705.3528 [hep-th]];
  Nucl.\ Phys.\  B {\bf 796} (2008) 457
  [arXiv:0710.0952 [hep-th]].

\bibitem{Damour:1987vm}
  T.~Damour and S.~Deser,
  Annales Poincare Phys.\ Theor.\  {\bf 47} (1987) 277.

\bibitem{Ouvry:1986dv}
  S.~Ouvry and J.~Stern,
  Phys.\ Lett.\  B {\bf 177} (1986) 335.

\bibitem{Siegel:1986zi}
  W.~Siegel and B.~Zwiebach,
  Nucl.\ Phys.\  B {\bf 282} (1987) 125.

\end{thebibliography}
\end{document}